\documentclass[10pt]{iopart}
\usepackage{iopams}  

\input{texdraw.sty}

\usepackage{graphicx}
\usepackage{pstricks}

\begin{document}

\topical[Cluster Variation Method]{Cluster Variation Method in
Statistical Physics and Probabilistic Graphical Models}

\author{Alessandro Pelizzola}

\address{Dipartimento di Fisica, Politecnico di Torino, c. Duca degli
  Abruzzi 24, 10129 Torino, Italy and INFN, Sezione di Torino} 

\begin{abstract}

The cluster variation method (CVM) is a hierarchy of approximate
variational techniques for discrete (Ising--like) models in
equilibrium statistical mechanics, improving on the mean--field
approximation and the Bethe--Peierls approximation, which can be
regarded as the lowest level of the CVM. In recent years it has been
applied both in statistical physics and to inference and optimization
problems formulated in terms of probabilistic graphical models. 

The foundations of the CVM are briefly reviewed, and the relations
with similar techniques are discussed. The main properties of the
method are considered, with emphasis on its exactness for particular
models and on its asymptotic properties.

The problem of the minimization of the variational free energy, which
arises in the CVM, is also addressed, and recent results about both
provably convergent and message-passing algorithms are discussed.

\end{abstract}


\submitto{\JPA}

\ead{alessandro.pelizzola@polito.it}

\maketitle

\section{Introduction}


The Cluster Variation Method (CVM) was introduced by Kikuchi
\cite{Kik51} in 1951, as an approximation technique for the
equilibrium statistical mechanics of lattice (Ising--like) models,
generalizing the Bethe--Peierls \cite{Bet35,Pei36} and
Kramers--Wannier \cite{KraWan1,KraWan2} approximations, an account of
which can be found in several textbooks \cite{PliBer,LavBel}. Apart
from rederiving these methods, Kikuchi proposed a combinatorial
derivation of what today we can call the cube (respectively triangle,
tetrahedron) approximation of the CVM for the Ising model on the
simple cubic (respectively triangular, face centered cubic) lattice. 

After the first proposal, many reformulations and applications, mainly
to the computation of phase diagram of lattice models in statistical
physics and material science, appeared, and have been reviewed in
\cite{PTPS}. The main line of activity has dealt with homogeneous,
translation--invariant lattice models with classical, discrete degrees
of freedom, but several other directions have been followed, including
for instance models with continuous degrees of freedom
\cite{KikuchiCont}, free surfaces \cite{MoranLopez,BuzPel}, models of
polymers \cite{Aguilera,LiseMaritan} and quantum models
\cite{MoritaQuant,Danani}. Out of equilibrium properties have also
been studied, in the framework of the path probability method
\cite{Ishii,Ducastelle,WadaKaburagi}, which is the dynamical version
of the CVM. Despite the CVM predicts mean--field like critical
behaviour, the problem of extracting critical behaviour from sequences
of CVM approximations has also been considered by means of different
approaches \cite{CVPAM1,CVPAM2,CVPAM3,CVPAM4,CAM}.

A line of research which is particularly relevant to the present
discussion has considered heterogeneous and random models. Much work
has been devoted in the 80's to applications of the CVM to models with
quenched random interactions (see e.g.\ \cite{SeinoKatsura} and refs.\
therein), mainly aiming to the phase diagram, and related equilibrium
properties, of Ising--like models of spin glasses in the average
case. The most common approach was based on the distribution of the
effective fields, and population dynamics algorithms were developed
and studied for the corresponding integral equations. All this effort
was however limited at the replica--symmetric level. Approaches taking
into account the first step of replica symmetry breaking have been
developed only recently \cite{SPScience}, at the level of the
Bethe--Peierls approximation, in its cavity method formulation, for
models on random graphs in both the single instance and average
case. These approaches have been particularly successful in their
application to combinatorial optimization problems, like
satisfiability \cite{SPSAT} and graph coloring \cite{SPCOL}. Another
interesting approach going in a similar direction has been proposed
recently \cite{Jort}, which relies on the analysis of the time
evolution of message--passing algorithms for the Bethe--Peierls
approximation.

Prompted by the interest in optimization and, more generally,
inference problems, a lot of work on the CVM has been done in recent
years also by researchers working on probabilistic graphical models
\cite{Smy97}, since the relation between the Bethe--Peierls
approximation and the belief propagation method \cite{Pearl} was
recognized \cite{Yed01}. The interaction between the two communities
of researchers working on statistical physics and optimization and
inference algorithms then led to the discovery of several new
algorithms for the CVM variational problem, and to a deeper
understanding of the method itself. There have been applications in
the fields of image restoration
\cite{TanMor,Tan02,Tanetal03,Tanetal04}, computer vision
\cite{FrePasCar}, interference in two--dimensional channels
\cite{Noam}, decoding of error--correcting codes
\cite{Gallager,McEliece,KabSaaLDPCC}, diagnosis \cite{Diagnosis},
unwrapping of phase images \cite{Unwrapping}, bioinformatics
\cite{BurgeKarlin,BioSeqAn,Krogh}, language processing
\cite{Huang,Manning}.

The purpose of the present paper is to give a short account of recent
advances on methodological aspects, and therefore applications will
not be considered in detail. It is not meant to be exhaustive and the
material included reflects in some way the interests of the
author. The plan of the paper is as follows. In \Sref{SMM-PGM} the
basic definitions for statistical mechanics and probabilistic
graphical models are given, and notation is established. In
\Sref{Fundamentals} the CVM is introduced in its modern formulation,
and in \Sref{RegionBased} it is compared with related approximation
techniques. Its properties are then discussed, with particular
emphasis on exact results, in \Sref{Exact}. Finally, the use of the CVM
as an approximation and the algorithms which can be used to solve the
CVM variational problem are illustrated in \Sref{Approx}. Conclusions
are drawn in \Sref{Conclusions}.

\section{Statistical mechanical models and probabilistic graphical
  models}
\label{SMM-PGM}

We are interested in dealing with models with discrete degrees of
freedom which will be denoted by $\bi{s} = \{ s_1, s_2, \ldots s_N
\}$. For instance, variables $s_i$ could take values in the set
$\{ 0,1 \}$ (binary variables), $\{ -1, +1 \}$ (Ising spins), or $\{ 1,
2, \ldots q \}$ (Potts variables).

Statistical mechanical models are defined through an energy function,
usually called Hamiltonian, $H = H(\bi{s})$, and the corresponding
probability distribution at thermal equilibrium is the Boltzmann
distribution
\begin{equation}
p(\bi{s}) = \frac{1}{Z} \exp\left[ - H(\bi{s}) \right],
\end{equation}
where the inverse temperature $\beta = (k_B T)^{-1}$ has been absorbed
into the Hamiltonian and
\begin{equation}
Z \equiv \exp(-F) = \sum_{\bi{s}} \exp\left[ - H(\bi{s}) \right]
\end{equation}
is the partition function, with $F$ the free energy.

\begin{figure}
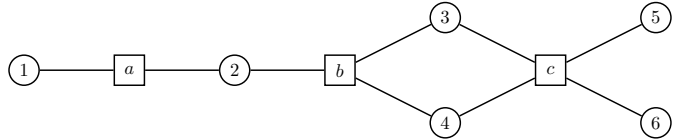

\begin{center}

\pspicture(-3,-1)(10,3)
\scalebox{0.7}{
\pscircle(0,1){.3}
\pscircle(4,1){.3}
\pscircle(8,0){.3}
\pscircle(8,2){.3}
\pscircle(12,0){.3}
\pscircle(12,2){.3}
\rput(0,1){1}
\rput(4,1){2}
\rput(8,0){4}
\rput(8,2){3}
\rput(12,0){6}
\rput(12,2){5}
\psframe(1.7,.7)(2.3,1.3)
\psframe(5.7,.7)(6.3,1.3)
\psframe(9.7,.7)(10.3,1.3)
\rput(2,1){$a$}
\rput(6,1){$b$}
\rput(10,1){$c$}
\psline(.3,1)(1.7,1)
\psline(2.3,1)(3.7,1)
\psline(4.3,1)(5.7,1)
\psline(6.3,1.15)(7.73,1.87)
\psline(6.3,.85)(7.73,.13)
\psline(8.27,1.87)(9.7,1.15)
\psline(8.27,.13)(9.7,.85)
\psline(10.3,1.15)(11.73,1.87)
\psline(10.3,.85)(11.73,.13)
}
\endpspicture

\end{center}
\caption{\label{FactorGraph}An example of a factor graph: variable and
  function nodes are denoted by circles and squares, respectively}
\end{figure}

The Hamiltonian is typically a sum of terms, each involving a small
number of variables. A useful representation is given by the {\it
factor graph} \cite{Kschischang}. A factor graph is a bipartite graph
made of variable nodes $i, j, \ldots$, one for each variable, and {\it
function nodes} $a, b, \ldots$, one for each term of the
Hamiltonian. An edge joins a variable node $i$ and a function node $a$
if and only if $i \in a$, that is the variable $s_i$ appears in $H_a$,
the term of the Hamiltonian associated to $a$. The Hamiltonian can
then be written as
\begin{equation}
H = \sum_a H_a(\bi{s_a}), \qquad \bi{s_a} = \{ s_i, i \in a \}.
\label{HsumHa}
\end{equation}
A simple example of a factor graph is reported in 
\Fref{FactorGraph}, and the corresponding Hamiltonian is written as
\begin{equation}
\fl H(s_1,s_2,s_3,s_4,s_5,s_6) = H_a(s_1,s_2) + H_b(s_2,s_3,s_4) +
H_c(s_3,s_4,s_5,s_6).
\end{equation}

The factor graph representation is particularly useful for models with
non--pairwise interactions. If the Hamiltonian contains only
1--variable and 2--variable terms, as in the Ising model
\begin{equation}
H = - \sum_i h_i s_i - \sum_{(i,j)} J_{ij} s_i s_j,
\label{Ising}
\end{equation}
then it is customary to draw a simpler graph, where only variable
nodes appear, and edges are drawn between pairs of interacting spins
$(i,j)$. In physical models the interaction strength $J_{ij}$ can
depend on the distance between spins, and interaction is often
restricted to nearest neighbours (NNs), which are denoted by $\langle
i,j \rangle$.

In combinatorial optimization problems, the Hamiltonian plays the role
of a cost function, and one is interested in the low--temperature
limit $T \to 0$, where only minimal energy states (ground states) have
a non--vanishing probability. 

Probabilistic graphical models \cite{Smy97,Lauritzen} are usually
defined in a slightly different way. In the case of {\it Markov random
fields}, also called {\it Markov networks}, the joint distribution over
all variables is given by
\begin{equation}
p(\bi{s}) = \frac{1}{Z} \prod_a \psi_a(\bi{s_a}),
\end{equation}
where $\psi_a$ is called {\it potential} (potentials involving only one
variable are often called {\it evidences}) and 
\begin{equation}
Z = \sum_{\bi{s}} \prod_a \psi_a(\bi{s_a}).
\end{equation}
Of course, a statistical mechanical model described by the Hamiltonian
(\ref{HsumHa}) corresponds to a probabilistic graphical models with
potentials $\psi_a = \exp(-H_a)$.  On the other hand, {\it Bayesian
networks}, which we will not consider here in detail, are defined in
terms of directed graphs and conditional probabilities. It must be
noted, however, that a Bayesian network can always be mapped onto a
Markov network \cite{Smy97}.

\section{Fundamentals of the Cluster Variation Method}
\label{Fundamentals}

The original proposal by Kikuchi \cite{Kik51} was based on an
approximation for the number of configurations of a lattice model with
assigned local expectation values.  The formalism was rather involved
to deal with in the general case, and since then many reformulations
came. A first important step was taken by Barker \cite{Bar53}, who
derived a computationally useful expression for the entropy
approximation. This was then rewritten as a cumulant expansion by
Morita \cite{Mor57,Mor72}, and Schlijper \cite{Sch83} noticed that
this expansion could have been written in terms of a M\"obius
inversion. A clear and simple formulation was then eventually set up
by An \cite{An88}, and this is the one we shall follow below.

The CVM can be derived from the
variational principle of equilibrium statistical mechanics, where the
free energy is given by
\begin{equation}
F = - \ln Z = \min_p {\cal F}(p) = \min_p \sum_{\bi{s}}
    \left[ p(\bi{s}) H(\bi{s}) + p(\bi{s}) \ln p(\bi{s}) \right]
\label{VarPrin}
\end{equation}
subject to the normalization constraint
\begin{equation}
\sum_{\bi{s}} p(\bi{s}) = 1.
\end{equation}
It is easily verified that the minimum is obtained for the Boltzmann
distribution 
\begin{equation}
\hat p(\bi{s}) = \frac{1}{Z} \exp[- H(\bi{s})] = {\rm arg} \,{\rm min}
\, {\cal F}
\end{equation}
and that the variational free energy can be written in the form of a
Kullback--Leibler distance
\begin{equation}
{\cal F}(p) = F + \sum_{\bi{s}} p(\bi{s}) \ln \frac{p(\bi{s})}{\hat
p(\bi{s})}.
\end{equation}

The basic idea underlying the CVM is to treat exactly the first term
(energy) of the variational free energy ${\cal F}(p)$ in
\Eref{VarPrin} and to approximate the second one (entropy) by means of
a truncated cumulant expansion. 

We first define a {\it cluster} $\alpha$ as a subset of the factor
graph such that if a factor node belongs to $\alpha$, then all the
variable nodes $i \in a$ also belong to $\alpha$ (while the converse
needs not to be true, otherwise the only legitimate clusters would
be the connected components of the factor graph). Given a cluster we
can define its energy
\begin{equation}
H_\alpha(\bi{s_\alpha}) = \sum_{a \in \alpha}  H_a(\bi{s_a}),
\end{equation}
probability distribution
\begin{equation}
p_\alpha(\bi{s_\alpha}) = \sum_{\bi{s} \setminus \bi{s_\alpha}} p(\bi{s})
\end{equation}
and entropy
\begin{equation}
S_\alpha = - \sum_{\bi{s_\alpha}} p_\alpha(\bi{s_\alpha}) \ln
p_\alpha(\bi{s_\alpha}). 
\end{equation}

Then the entropy cumulants are defined by
\begin{equation}
S_\alpha = \sum_{\beta \subseteq \alpha} \tilde S_\beta,
\end{equation}
which can be solved with respect to the cumulants by means of a
M\"obius inversion, which yields
\begin{equation}
\tilde S_\beta = \sum_{\alpha \subseteq \beta}
(-1)^{n_\alpha - n_\beta} S_\alpha,
\end{equation}
where $n_\alpha$ denotes the number of variables in cluster
$\alpha$. The variational free energy can then be written as
\begin{equation}
{\cal F}(p) = \sum_{\bi{s}} p(\bi{s}) H(\bi{s}) - \sum_\beta \tilde
S_\beta, 
\end{equation}
where the second summation is over all possible clusters. 

The above equation is still an exact one, and here the approximation
enters. A set $R$ of clusters, made of maximal clusters and all their
subclusters, is selected, and the cumulant expansion of the entropy is
truncated retaining only terms corresponding to clusters in $R$. In
order to treat the energy term exactly it is necessary that each
function node is contained in at least one maximal cluster. One gets
\begin{equation}
\sum_\beta \tilde S_\beta \simeq \sum_{\beta \in R} \tilde S_\beta =
\sum_{\alpha \in R} a_\alpha S_\alpha,
\label{CVMapprox}
\end{equation}
where the coefficients $a_\alpha$, sometimes called M\"obius numbers,
satisfy \cite{An88}
\begin{equation}
\sum_{\beta \subseteq \alpha \in R} a_\alpha = 1 \qquad
  \forall \beta \in R.
\label{MobiusNumbers}
\end{equation}
The above condition means that every subcluster must be counted
exactly once in the entropy expansion and allows to rewrite also the
energy term as a sum of cluster energies, yielding the approximate
variational free energy
\begin{equation}
{\cal F}(\{p_\alpha, \alpha \in R\}) = \sum_{\alpha \in R} a_\alpha
{\cal F}_\alpha(p_\alpha),
\label{CVMFree}
\end{equation}
where the cluster free energies are given by
\begin{equation}
{\cal F}_\alpha(p_\alpha) = \sum_{\bi{s_\alpha}} \left[
  p_\alpha(\bi{s_\alpha)} H_\alpha(\bi{s_\alpha}) + 
  p_\alpha(\bi{s_\alpha)} \ln p_\alpha(\bi{s_\alpha)} \right].
\label{ClusterFree}
\end{equation}
The CVM then amounts to the minimization of the above variational free
energy with respect to the cluster probability distributions, subject
to the normalization
\begin{equation}
\sum_{\bi{s_\alpha}} p_\alpha(\bi{s_\alpha}) = 1 \qquad \forall \alpha
  \in R 
\end{equation}
and compatibility constraints
\begin{equation}
p_\beta(\bi{s_\beta)} = \sum_{\bi{s_{\alpha \setminus \beta}}}
  p_\alpha(\bi{s_\alpha}) \qquad  
  \forall \beta \subset \alpha \in R.
\label{CompConstr}
\end{equation}
It is of great importance to observe that the above constraint set is
approximate, in the sense that there are sets of cluster probability
distributions that satisfy these constraints and nevertheless cannot
be obtained as marginals of a joint probability distribution. An
explicit example will be given in \Sref{Exact}.

The simplest example is the pair approximation for a model with
pairwise interactions, like the Ising model (\ref{Ising}). The maximal
clusters are the pairs of interacting variables, and the other
clusters appearing in $R$ are the variable nodes. The pairs have
M\"obius number 1, while for the variable nodes $a_i = 1 - d_i$, where
$d_i$ is the {\it degree} of node $i$, that is, in the factor graph
representation, the number of function nodes it belongs to. 

\begin{figure}
\begin{center}
\includegraphics*[scale=.5]{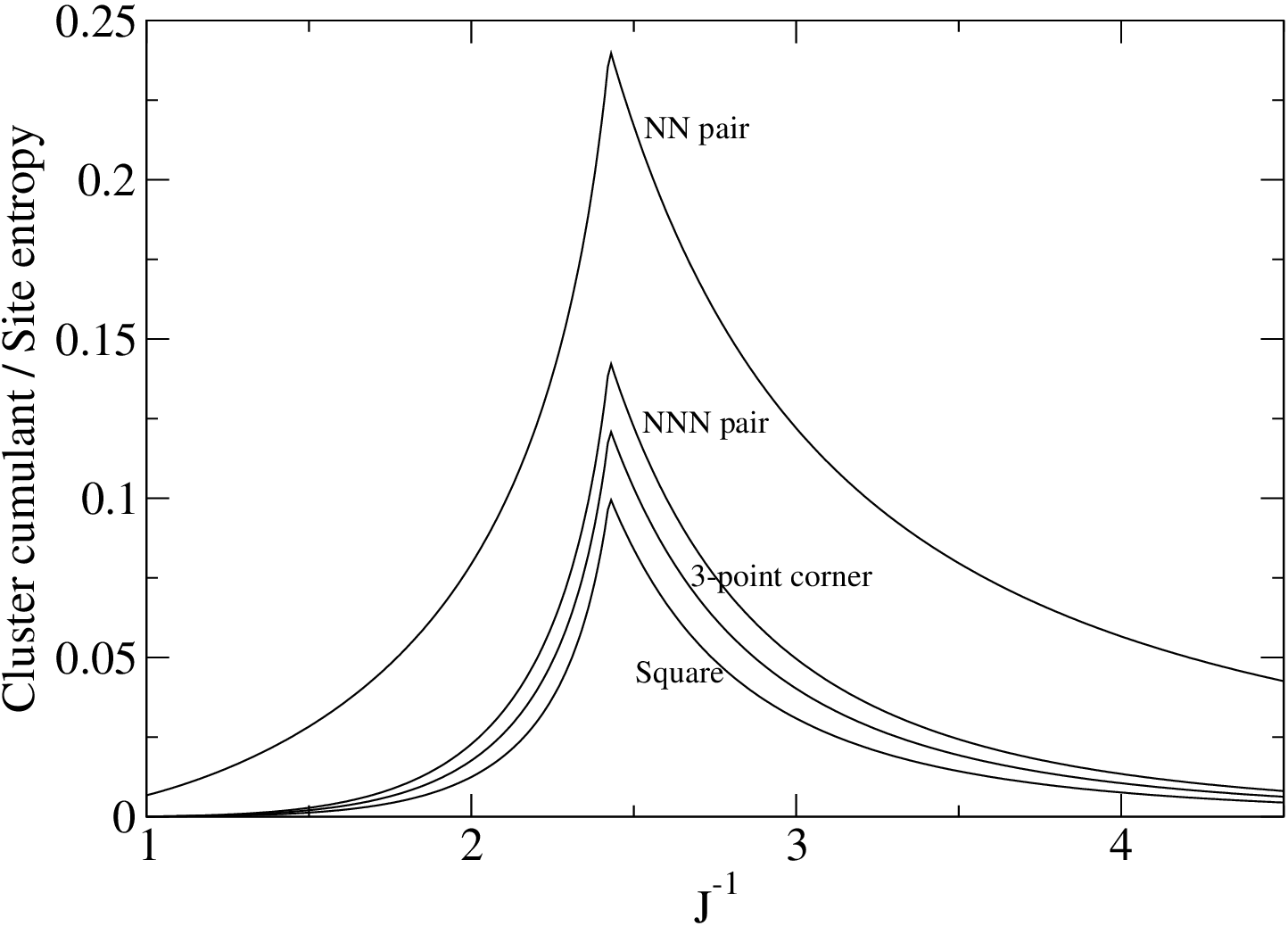}
\end{center}
\caption{\label{Cumulants}Cumulants for the square lattice Ising model}
\end{figure}

The quality of the approximation (\ref{CVMapprox}) depends on the value
of the neglected cumulants. In the applications to lattice systems it
is typically assumed that, since cumulants are related to
correlations, they vanish quickly for clusters larger than the
correlation length of the model. In \Fref{Cumulants} the first
cumulants, relative to the site (single variable) entropy, are shown
for the homogeneous ($J_{ij} = J$), zero field ($h_i = 0$), square
lattice Ising model, in the square approximation of the CVM.

It can be seen that the cumulants peak at the (approximate) critical
point and decrease as the cluster size increases. This property is not
however completely general, it may depend on the interaction range. It
has been shown \cite{KappenWiegerinck} that this does not hold for
finite instances of the Sherrington--Kirkpatrick spin--glass model,
which is a fully connected model.

The meaning of cumulants as a measure of correlation can be easily
understood by considering a pair of weakly correlated variables and
writing their joint distribution as
\begin{equation}
p_{12}(s_1,s_2) = p_1(s_1) p_2(s_2) \left[ 1 +
  \varepsilon \, q(s_1,s_2) \right], \qquad \varepsilon \ll 1.
\end{equation}
The corresponding cumulant is then
\begin{equation}
\tilde S_{12} = S_{12} - S_1 - S_2 = - \langle \ln \left[ 1 +
  \varepsilon \, q(s_1,s_2) \right] \rangle = \Or(\varepsilon).
\end{equation}

\section{Region--based free energy approximations}
\label{RegionBased}

The idea of {\it region--based free energy approximations}, put
forward by Yedidia \cite{Yed04}, is quite useful to elucidate some of
the characteristics of the method, and its relations to other
techniques. A region--based free energy approximation is formally
similar to the CVM, and can be defined through equations (\ref{CVMFree})
and (\ref{ClusterFree}), but the requirements on the coefficients
$a_\alpha$ are weaker. The single counting condition is imposed only
on variable and function nodes, instead of all subclusters:
\begin{eqnarray}
\sum_{\alpha \in R, a \in \alpha} a_\alpha = 1 \qquad \forall a, \\
\sum_{\alpha \in R, i \in \alpha} a_\alpha = 1 \qquad \forall i.
\end{eqnarray}

Interesting particular cases are obtained if $R$ contains only two
types of regions, {\it large regions} and {\it small regions}. The
{\it junction graph} method \cite{Yed04,AjiMc} is obtained if they
form a directed graph, with edges from large to small regions, such
that:
\begin{enumerate}
\item every edge connects a large region with a small region which is
a subset of the former; 
\item the subgraph of the regions containing a given node is a
connected tree.
\end{enumerate}
On the other hand, the {\it Bethe--Peierls approximation}, in its most general
formulation, is obtained by taking function nodes (with the associated
variable nodes) as large regions and variable nodes as small
regions. This reduces to the usual statistical physics formulation in
the case of pairwise interactions. 

The CVM is a special region--based free energy approximation, with the
property that $R$ is closed under intersection. Indeed, one could
define $R$ for the CVM as the set made of the maximal clusters and all
the clusters which can be obtained by taking all the possible
intersections of (any number of) maximal clusters. 

\begin{figure}
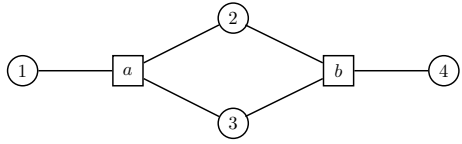

\begin{center}
\pspicture(-1,-2)(7,2)
\scalebox{0.7}{
\pscircle(0,0){.3}
\pscircle(4,1){.3}
\pscircle(4,-1){.3}
\pscircle(8,0){.3}
\psframe(1.7,-.3)(2.3,.3)
\psframe(5.7,-.3)(6.3,.3)
\rput(0,0){1}
\rput(4,1){2}
\rput(4,-1){3}
\rput(8,0){4}
\rput(2,0){$a$}
\rput(6,0){$b$}
\psline(.3,0)(1.7,0)
\psline(2.3,.15)(3.73,.87)
\psline(2.3,-.15)(3.73,-.87)
\psline(4.23,.87)(5.7,.15)
\psline(4.23,-.87)(5.7,-.15)
\psline(6.3,0)(7.7,0)
}
\endpspicture
\end{center}
\caption{\label{BetheNotCVM}Factor graph of a model for which the
  Bethe--Peierls approximation is not a special case of the CVM}
\end{figure}

It is easy to verify that the Bethe--Peierls approximation is a
special case of CVM only if no function node shares more than one
variable node with another function node. If this is not the case, one
should be careful when applying the Bethe--Peierls
approximation. Consider a model with the factor graph depicted in
\Fref{BetheNotCVM}, where $s_i = \pm 1$ ($i = 1, 2, 3, 4$), $H = H_a +
H_b$ and
\begin{eqnarray}
H_a(s_1,s_2,s_3) = - h_0 s_1 - \frac{h}{2} (s_2 + s_3) - J s_1 s_2 s_3,
 \\
H_b(s_2,s_3,s_4) = - h_0 s_4 - \frac{h}{2} (s_2 + s_3) - J s_2 s_3 s_4.
\end{eqnarray}

\begin{figure}
\begin{center}
\includegraphics*[scale=.38]{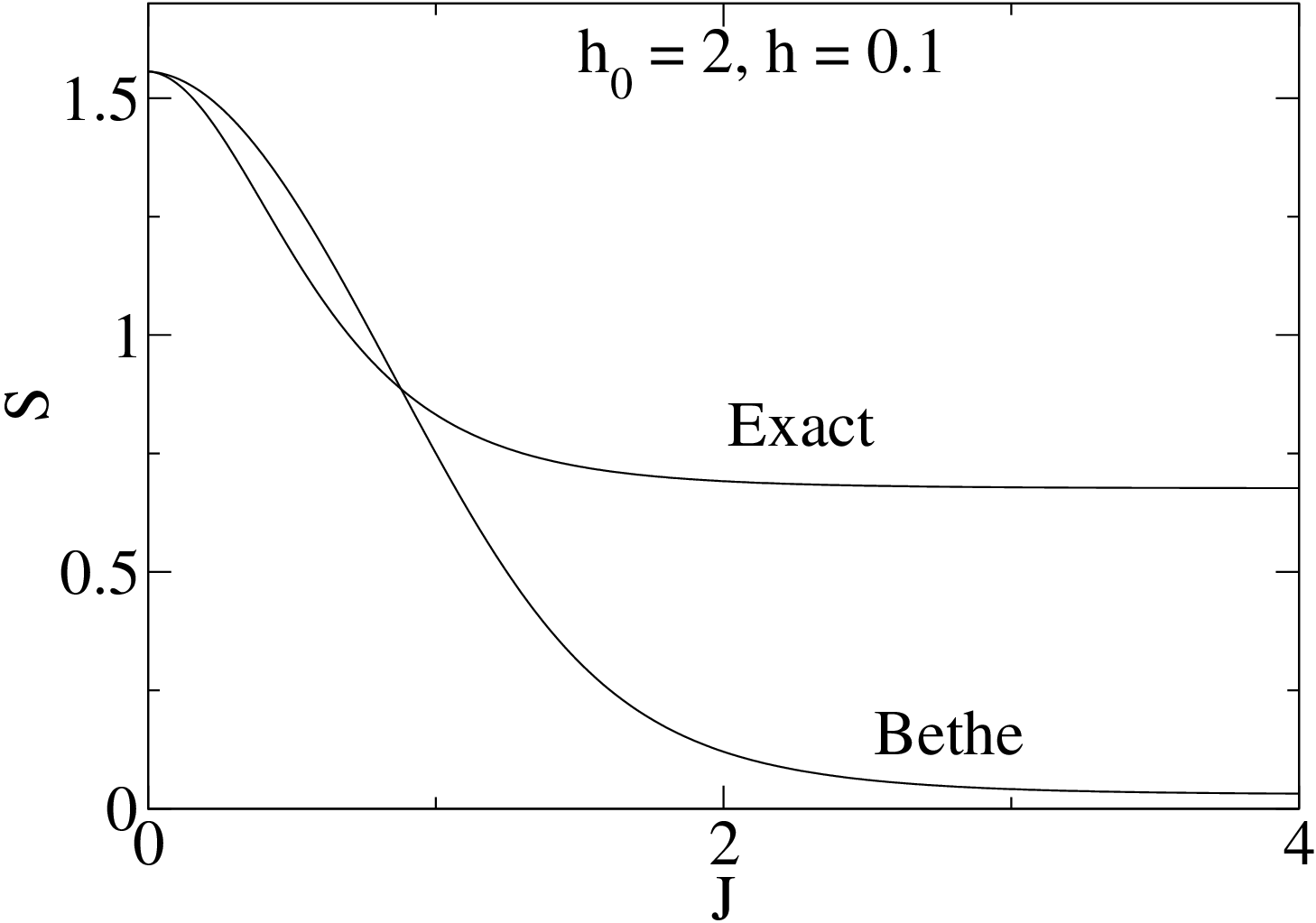}
\end{center}
\caption{\label{BetheFailure}Entropy of the Bethe--Peierls approximation vs the
exact one for a model for which the Bethe--Peierls approximation is not a
special case of the CVM}
\end{figure}

The CVM, with function nodes as maximal clusters, is exact (notice
that it coincides with the junction graph method), and the corresponding exact
cumulant expansion for the entropy is
\begin{equation}
S = S_a + S_b - S_{23},
\end{equation}
while the Bethe--Peierls entropy is 
\begin{equation}
S_{\rm BP} = S_a + S_b - S_2 - S_3.
\end{equation}
The two entropies differ by the cumulant $\tilde S_{23} = S_{23} - S_2
- S_3$, and hence correlations between variable nodes 2 and 3 cannot
be captured by the Bethe--Peierls approximation.  In
\Fref{BetheFailure} it is clearly illustrated how the Bethe--Peierls
approximation can fail. At large enough $J$ the exact entropy is
larger (by roughly $\ln 2$) than the Bethe--Peierls one.

\section{Exactly solvable cases}
\label{Exact}

The CVM is known to be exact in several cases, due to the topology of
the underlying graph, or to the special form of the Hamiltonian. In
the present section we shall first consider cases in which the CVM is
exact due to the graph topology, then proceed to the issue of
realizability and consider cases where the form of the Hamiltonian
makes an exact solution feasible with the CVM. 

\subsection{Tree-like graphs}

It is well known that the CVM is exact for models defined on
tree--like graphs. This statement can be made more precise by
referring to the concept of {\it junction tree} \cite{LauSpi,Jensen},
which we shall actually use in its generalized form given by Yedidia,
Freeman and Weiss \cite{Yed04}. A junction tree is a tree--like
junction graph. The corresponding large regions are often called {\it
cliques}, and the small regions {\it separators}. With reference to
\Fref{BetheNotCVM} it is easy to check that the CVM, as described in
the previous section, corresponds to a junction tree with cliques
$(a123)$ and $(b234)$ and separator $(23)$, while the junction graph
corresponding to the Bethe--Peierls approximation is not a tree. 

For a model defined on a junction tree the joint probability
distribution factors \cite{Yed04,Cowell} according to 
\begin{equation}
p(\bi{s}) = \frac{\displaystyle\prod_{\alpha \in R_L}
  p_{\alpha}(\bi{s_\alpha})} 
{\displaystyle\prod_{\beta \in R_S} p_\beta^{d_\beta-1}(\bi{s_\beta})},
\end{equation}
where $R_L$ and $R_S$ denote the sets of large and small regions,
respectively, and $d_\beta$ is the degree of the small region $\beta$
in the junction tree. Notice that no normalization is needed.

The above factorization of the probability leads to the exact cumulant
expansion 
\begin{equation}
S = \sum_{\alpha \in R_L} S_\alpha - \sum_{\beta \in R_S} (d_\beta-1)
S_\beta,
\end{equation}
and therefore the CVM with $R = R_L \cup R_S$ is exact.

\begin{figure}
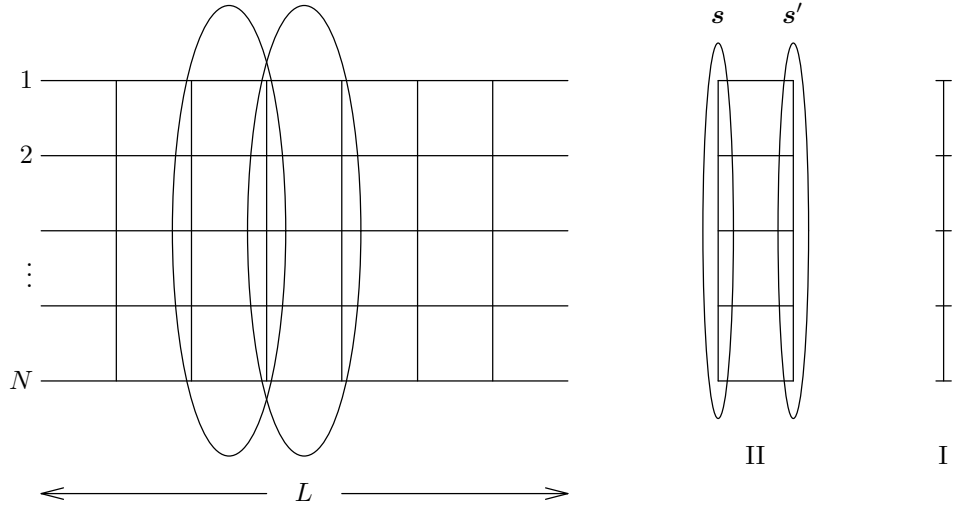

\centertexdraw{
\drawdim cm \linewd 0.02 
\arrowheadsize l:0.3 w:0.15
\arrowheadtype t:V
\move(0 0) \lvec(7 0) 
\move(0 1) \lvec(7 1) 
\move(0 2) \lvec(7 2) 
\move(0 3) \lvec(7 3) 
\move(0 4) \lvec(7 4)
\move(1 0) \lvec(1 4) 
\move(2 0) \lvec(2 4) 
\move(3 0) \lvec(3 4) 
\move(4 0) \lvec(4 4) 
\move(5 0) \lvec(5 4) 
\move(6 0) \lvec(6 4) 
\textref h:R v:C \htext(-.1 4) {1}
\textref h:R v:C \htext(-.1 3) {2}
\textref h:R v:C \htext(-.1 1.5) {$\vdots$}
\textref h:R v:C \htext(-.1 0) {$N$}
\move(2.5 2) \lellip rx:.75 ry:3
\move(3.5 2) \lellip rx:.75 ry:3
\textref h:C v:C \htext(3.5 -1.5) {$L$}
\move(3 -1.5) \avec(0 -1.5) \move(4 -1.5) \avec(7 -1.5)
\move(9 0) \lvec(9 4)
\move(10 0) \lvec(10 4)
\move(9 0) \lvec(10 0)
\move(9 1) \lvec(10 1)
\move(9 2) \lvec(10 2)
\move(9 3) \lvec(10 3)
\move(9 4) \lvec(10 4)
\move(9 2) \lellip rx:.2 ry:2.5
\textref h:C v:B \htext(9 4.75) {$\bi{s}$}
\textref h:C v:B \htext(10 4.75) {$\bi{s^\prime}$}
\move(10 2) \lellip rx:.2 ry:2.5
\textref h:C v:C \htext(9.5 -1) {II}
\move(12 0) \lvec(12 4)
\move(11.9 0) \lvec(12.1 0)
\move(11.9 1) \lvec(12.1 1)
\move(11.9 2) \lvec(12.1 2)
\move(11.9 3) \lvec(12.1 3)
\move(11.9 4) \lvec(12.1 4)
\textref h:C v:C \htext(12 -1) {I}
}
\caption{\label{Strip}A one--dimensional strip and the clusters used
  to solve a pairwise model on it}
\end{figure}

As a first example, consider a particular subset of the square
lattice, the strip depicted in \Fref{Strip}, with open boundary
conditions in the horizontal direction, and define on it a model with
pairwise interactions (we do not use the factor graph representation
here).

According to the junction tree property, the joint probability factors
as follows:
\begin{equation}
p(\bi{s}) = \frac{\displaystyle\prod_{\alpha \in {\rm II}}
  p_\alpha(\bi{s_\alpha})} 
{\displaystyle\prod_{\beta \in {\rm I}} p_\beta(\bi{s_\beta})}, 
\end{equation}
where I and II denote the sets of chains (except boundary ones) and
ladders, respectively, shown in \Fref{Strip}. As a consequence, the
cumulant expansion
\begin{equation}
S = \sum_{\alpha \in {\rm II}} S_\alpha - 
\sum_{\beta \in {\rm I}} S_\beta
\end{equation}
of the entropy is also exact, and the cluster variation method with $R
= {\rm II} \cup {\rm I}$ is exact. For strip width $N = 1$ we obtain
the well--known statement that the Bethe--Peierls approximation is
exact for a one--dimensional chain. Rigorous proofs of this statement
have been given by Brascamp \cite{Bra71} and Percus \cite{Per77}. More
generally, Schlijper has shown \cite{Sch84} that the equilibrium
probability of a $d$--dimensional statistical mechanical model with
finite range interactions is completely determined by its restrictions
(marginals) to $d-1$--dimensional slices of width at least equal to
the interaction range.

In the infinite length limit $L \to \infty$ translational invariance
is recovered
\begin{equation}
\fl \frac{{\cal F}}{L} = \sum_{\bi{s},\bi{s^\prime}} \left([ p_{\rm
II}(\bi{s},\bi{s^\prime}) H_{\rm II}(\bi{s},\bi{s^\prime}) + p_{\rm
II}(\bi{s},\bi{s^\prime}) \ln p_{\rm II}(\bi{s},\bi{s^\prime})\right]
- \sum_{\bi{s}} p_{\rm I}(\bi{s}) \ln p_{\rm I}(\bi{s})
\end{equation}
and solving for $p_{\rm II}$ we obtain the transfer matrix formalism:
\begin{eqnarray}
\frac{F}{L} = - \ln \max_{p_{\rm I}} \left\{
\sum_{\bi{s},\bi{s^\prime}} 
p_{\rm I}^{1/2}(\bi{s}) \exp\left[ -
H_{\rm II}(\bi{s},\bi{s^\prime}) \right]
p_{\rm I}^{1/2}(\bi{s^\prime}) \right\} \\
\sum_{\bi{s}} p_{\rm I}(\bi{s}) = 1
\end{eqnarray}

The natural iteration method (see \sref{VarAlg}) in this case reduces to
the power method for finding the largest eigenvalue of the transfer
matrix.

\begin{figure}
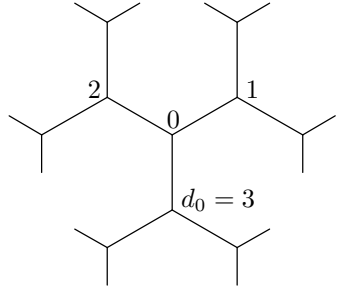

\centertexdraw{
\drawdim cm \linewd 0.02 
\arrowheadsize l:0.3 w:0.15
\arrowheadtype t:V
\move(0 0) \lvec(.866 .5)
\move(.866 .5) \lvec(1.732 0)
\move(1.732 0) \lvec(1.732 -.5)
\move(1.732 0) \lvec(2.165 .25)
\move(.866 .5) \lvec(.866 1.5)
\move(.866 1.5) \lvec(1.299 1.75)
\move(.866 1.5) \lvec(.433 1.75)
\move(0 0) \lvec(-.866 .5)
\move(-.866 .5) \lvec(-1.732 0)
\move(-1.732 0) \lvec(-1.732 -.5)
\move(-1.732 0) \lvec(-2.165 .25)
\move(-.866 .5) \lvec(-.866 1.5)
\move(-.866 1.5) \lvec(-1.299 1.75)
\move(-.866 1.5) \lvec(-.433 1.75)
\move(0 0) \lvec(0 -1)
\move(0 -1) \lvec(.866 -1.5)
\move(.866 -1.5) \lvec(.866 -2)
\move(.866 -1.5) \lvec(1.299 -1.25)
\move(0 -1) \lvec(-.866 -1.5)
\move(-.866 -1.5) \lvec(-.866 -2)
\move(-.866 -1.5) \lvec(-1.299 -1.25)
\textref h:C v:B \htext(0 .1) {0}
\textref h:L v:B \htext(.966 .5) {1}
\textref h:R v:B \htext(-.966 .5) {2}
\textref h:L v:B \htext(.1 -1) {$d_0 = 3$}
}
\caption{\label{BetheLattice}A small portion of a tree}
\end{figure}

As a second example, consider a tree, like the one depicted in
\Fref{BetheLattice}, and a model with pairwise interactions defined on
it.

In this case the probability factors according to
\begin{equation}
p(\bi{s}) = \frac{\displaystyle\prod_{\langle i j \rangle} p_{ij}(s_i,s_j)}
{\displaystyle\prod_{i} p_i^{d_i-1}(s_i)},
\end{equation}
where $\langle i j \rangle$ denotes a pair of adjacent nodes. The
cumulant expansion of the entropy is therefore
\begin{equation}
S = \sum_{\langle i j \rangle} S_{ij} - \sum_{i} (d_i - 1) S_i,
\end{equation}
and the pair approximation of the CVM (coinciding with Bethe--Peierls and
junction graph) is exact. Recently this property has been exploited to
study models on finite connectivity random graphs, which strictly
speaking are not tree--like: loops are present, but in the
thermodynamic limit their typical length scales like $\ln N$ \cite{Bollobas}.

\begin{figure}
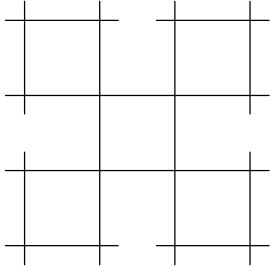

\centertexdraw{
\drawdim cm \linewd 0.02 
\arrowheadsize l:0.3 w:0.15
\arrowheadtype t:V
\move(-1.75 -1.5) \lvec(-.25 -1.5)
\move(1.75 -1.5) \lvec(.25 -1.5)
\move(-1.75 -.5) \lvec(1.75 -.5)
\move(-1.75 .5) \lvec(1.75 .5)
\move(-1.75 1.5) \lvec(-.25 1.5)
\move(1.75 1.5) \lvec(.25 1.5)

\move(-1.5 -1.75) \lvec(-1.5 -.25)
\move(-1.5 1.75) \lvec(-1.5 .25)
\move(-.5 -1.75) \lvec(-.5 1.75)
\move(.5 -1.75) \lvec(.5 1.75)
\move(1.5 -1.75) \lvec(1.5 -.25)
\move(1.5 1.75) \lvec(1.5 .25)
}
\caption{\label{Cactus}A small portion of a square cactus lattice}
\end{figure}

As a final example, consider the so--called (square) cactus lattice
(the interior of a Husimi tree), depicted in \Fref{Cactus}.

Here the probability factors according to
\begin{equation}
p(\bi{s}) = \frac{\displaystyle\prod_{\opensquare}
p_{\opensquare}(\bi{s_{\opensquare}})}{\displaystyle\prod_i p_i(s_i)},
\end{equation}
the entropy cumulant expansion takes the form
\begin{equation}
S = \sum_{\opensquare} S_{\opensquare} - \sum_{i} S_i,
\end{equation}
and the CVM with $R$ made of square plaquettes and sites is
exact. Again, this coincides with the junction graph method and, if
function nodes are associated to square plaquettes (so that the
corresponding factor graph is tree--like), with Bethe--Peierls. 

\subsection{Realizability}

We have seen that when the probability factors in a suitable way, the
CVM can be used to find an exact solution. By analogy, we could ask
whether, as in the mean field approximation, CVM approximations can
yield an estimate of the joint probability distribution as a function
of the cluster distributions, in a factorized form. In the general
case, the answer is negative. One cannot, using a trial factorized
form like
\begin{equation}
\prod_\alpha [ p_\alpha(\bi{s_\alpha}) ]^{a_\alpha}
\label{CVMproduct}
\end{equation}
(which would lead to a free energy like that in Eqs.\
\ref{CVMFree}-\ref{ClusterFree}), obtain a joint probability
distribution which marginalizes down to the cluster probability
distributions used as a starting point. As a consequence, we have no
guarantee that the CVM free energy is an upper bound to the exact free
energy. Moreover, in sufficiently frustrated problems, the cluster
probability distributions cannot even be regarded as marginals of a
joint probability distribution \cite{Sch88}.

It can be easily checked that \Eref{CVMproduct} is not, in the general
case, a probability distribution. It is not normalized and therefore
its marginals do not coincide with the $p_\alpha$'s used to build
it. At best, one can show that
\begin{equation}
\prod_\alpha [ p_\alpha(\bi{s_\alpha}) ]^{a_\alpha} \propto
\exp[-H(\bi{s})],
\label{FactorProp}
\end{equation}
but the normalization constant is unknown. This has been proven in
\cite{WaiJaaWil} at the Bethe--Peierls level, and the proof can be
easily generalized to any CVM approximation.

Let us now focus on a very simple example. Consider three Ising
variables, $s_i = \pm 1$, $i = 1, 2, 3$, with the following node and
pair probabilities:
\begin{eqnarray}
p_i(s_i) = 1/2 \qquad i = 1, 2, 3 \\ 
p_{ij}(s_i,s_j) = \frac{1 + c s_i
s_j}{4}, \qquad -1 \le c \le 1, \qquad i < j.
\end{eqnarray}
A joint $p(s_1,s_2,s_3)$ marginalizing to the above probabilities
exists for $-1/3 \le c \le 1$, which shows clearly that the constraint
set \Eref{CompConstr} is approximate, and in particular it can be too
loose. For instance, in \cite{PelPre} it has been shown that due to
this problem the Bethe--Peierls approximation for the triangular Ising
antiferromagnet predicts, at low temperature, unphysical results for
the correlations and a negative entropy.

Moreover, the joint probability $p(s_1,s_2,s_3)$ is given by the
CVM--like factorized form
\begin{equation}
\frac{p_{12}(s_1,s_2) p_{13}(s_1,s_3) p_{23}(s_2,s_3)}{p_1(s_1)
  p_2(s_2) p_3(s_3)} 
\end{equation}
only if $c = 0$, that is if the variables are completely uncorrelated.

As a more interesting case, we shall consider in the next subsection
the square lattice Ising model. In this case it has been shown
\cite{Disorder1,Disorder2} that requiring realizability yields an
exactly solvable case.

\subsection{Disorder points}

For a homogeneous (translation--invariant) model defined on a
square lattice, the square approximation of the CVM, that is
the approximation obtained by taking the elementary square plaquettes
as maximal clusters, entails the following approximate entropy
expansion:
\begin{equation}
S \simeq \sum_{\opensquare} S_{\opensquare} - \sum_{\langle i j \rangle}
S_{ij} + \sum_i S_i.
\label{SquareEntropy}
\end{equation}
The corresponding factorization 
\begin{equation}
\prod_{\opensquare}
  p_{\opensquare}(\bi{s_{\opensquare}}) \prod_{\langle i j \rangle}
  p_{ij}^{-1}(s_i,s_j) \prod_i p_i(s_i) 
\label{pDisorder}
\end{equation}
for the probability does not, in general, give an approximation to the
exact equilibrium distribution. Indeed, it does not marginalize to the
cluster distributions and is not even normalized. 

\begin{figure}
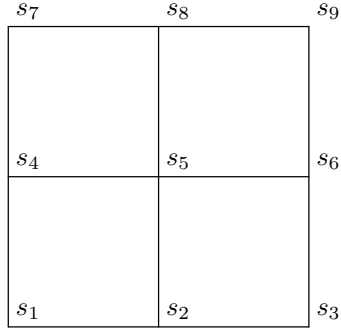

\centertexdraw{
\drawdim cm \linewd 0.02 
\arrowheadsize l:0.3 w:0.15
\arrowheadtype t:V
\move(0 0) \lvec(4 0) 
\move(0 2) \lvec(4 2) 
\move(0 4) \lvec(4 4)
\move(0 0) \lvec(0 4) 
\move(2 0) \lvec(2 4) 
\move(4 0) \lvec(4 4) 
\textref h:L v:B
\htext(.1 .1) {$s_1$}
\htext(2.1 .1) {$s_2$}
\htext(4.1 .1) {$s_3$}
\htext(.1 2.1) {$s_4$}
\htext(2.1 2.1) {$s_5$}
\htext(4.1 2.1) {$s_6$}
\htext(.1 4.1) {$s_7$}
\htext(2.1 4.1) {$s_8$}
\htext(4.1 4.1) {$s_9$}
}
\caption{\label{Square3x3}A $3 \times 3$ square on the square lattice}
\end{figure}

One could, however, try to impose that the joint probability given by
the above factorization marginalizes to the cluster
distributions. It turns out that it is sufficient to impose
such a condition on the probability distribution of a $3 \times 3$
square, like the one depicted in \Fref{Square3x3}. It is easy to check
that for an Ising model the CVM--like function
\begin{equation}
\fl
\frac{
p_{4}(s_1,s_2,s_5,s_4)
p_{4}(s_2,s_3,s_6,s_5)
p_{4}(s_4,s_5,s_8,s_7)
p_{4}(s_5,s_6,s_9,s_8)
p_{1}(s_5)}
{p_{2}(s_2,s_5) p_{2}(s_5,s_8)
p_{2}(s_4,s_5) p_{2}(s_5,s_6)}
\end{equation}
marginalizes to the square, pair and site distributions ($p_4$, $p_2$
and $p_1$ respectively) only if odd expectation values vanish and
\begin{equation}
\langle s_i s_k \rangle_{\langle \langle i k \rangle \rangle} =
  \langle s_i s_j \rangle_{\langle i j \rangle}^2,
\end{equation}
where the l.h.s.\ is the next nearest neighbour correlation, while the
r.h.s.\ is the square of the nearest neighbour correlation.

Leaving apart the trivial non--interacting case, the above condition
is satisfied by an Ising model with nearest neighbour, next nearest
neighbour and plaquette interactions, described by the Hamiltonian 
\begin{equation}
H = - J_1 \sum_{\langle i j \rangle} s_i s_j 
- J_2 \sum_{\langle \langle i j \rangle \rangle} s_i s_j
- J_4 \sum_{\opensquare} s_i s_j s_k s_l,
\end{equation}
if the couplings satisfy the {\it disorder} condition (see
\cite{Disorder1} and refs.\ therein)
\begin{equation}
\cosh (2 J_1) = \frac{e^{4J_2+2J_4}+e^{-4J_2+2J_4}+2 e^{-2J_2}}
{2\left(e^{2J_2}+e^{2J_4}\right)}.
\end{equation}
This defines a variety in the parameter space, lying in the disordered
phase of the model, and in particular in the region where nearest
neighbour and next nearest neighbour interactions compete.  In this case
the square approximation of the CVM yields the exact solution,
including the exact free energy density
\begin{equation}
f = - \ln \left[ \exp(-J_4)+\exp(J_4 - 2J_2) \right],
\end{equation}
and the nearest neighbour correlation
\begin{equation}
g = \langle s_i s_j \rangle_{\langle i j \rangle} = 
\frac{\exp(-4J_2) - \cosh(2 J_1)}{\sinh(2J_1)}.
\end{equation}
Higher order correlations can be derived from the joint probability
\Eref{pDisorder}, for example the two--body correlation function
$\Gamma(x,y) = \langle s(x_0,y_0) s(x_0+x,y_0+y) \rangle$ (where spin
variables have been identified by their coordinates on the lattice),
which simply reduces to a power of the nearest neighbour correlation:
$\Gamma(x,y) = g^{|x|+|y|}$. For this reason a line of disorder points
is often referred to as a one--dimensional line. Or the plaquette
correlation:
\begin{equation}
q = \langle s_i s_j s_k s_l \rangle_{\opensquare} = 
\frac{e^{4J_4}\left(1-e^{8J_2}\right) +
4 e^{2J_2}\left(e^{2J_4}-e^{2J_2}\right)} 
{e^{4J_4}\left(1-e^{8J_2}\right) +
4 e^{2J_2}\left(e^{2J_4}+e^{2J_2}\right)}.
\end{equation}
Finally, since all the pair correlations are given simply as powers of
the nearest--neighbour correlation we can easily calculate the
momentum space correlation function, or structure factor. We first
write $\Gamma(x,y) =
\exp\left(-\displaystyle\frac{|x|+|y|}{\xi}\right)$, where $\xi =
-(\ln g)^{-1}$. After a Fourier transform one finds $S(p_x,p_y) =
S_1(p_x) S_1(p_y)$, where
\begin{equation}
S_1(p) = \frac{\sinh(1/\xi)}{\cosh(1/\xi) - \cos p}.
\end{equation}
It can be verified that the structure factors calculated by Sanchez
\cite{Sanchez} and (except for a misprint) Cirillo and coworkers
\cite{Cirillo} reduce to the above expression on the disorder line.

\subsection{Wako--Sait\^o--Mu\~noz--Eaton model of protein folding}

There is at least another case in which the probability factors at the
level of square plaquettes, and the CVM yields the exact solution. It
is the Wako--Sait\^o--Mu\~noz--Eaton model of protein folding
\cite{WakSat1,WakSat2,MunEat1,MunEat2,MunEat3,BruPel1,BruPel2,PelJSTAT}. Here
we will not delve into the details of the model, giving only its
Hamiltonian in the form
\begin{equation}
H = \sum_{i=1}^L \sum_{j=i}^L h_{i,j} x_{i,j}, \qquad x_{i,j} =
\prod_{k=i}^j x_k, \qquad x_k = 0, 1.
\end{equation}
It is a one--dimensional model with arbitrary range multivariable
interactions, but the particular form of these interactions makes an
exact solution possible. A crucial step in this solution was the
mapping to a two--dimensional model \cite{BruPel1}, where the
statistical variables are the $x_{i,j}$'s (see \Fref{MunozEaton} for
an illustration). In terms of these variables the Hamiltonian is
local, and the only price one has to pay is to take into account the
constraints
\begin{equation}
x_{i,j} = x_{i+1,j} x_{i,j-1},
\end{equation}
which can be viewed as local interactions. 

\begin{figure}
\begin{center}
\psset{unit=.7cm}
\pspicture(-2,-11)(11,2)
\psline(-1,-.5)(11,-.5)
\psline(.5,-11)(.5,1)
\rput(1,0){1}
\rput(2,0){2}
\rput(3,0){3}
\rput(4,0){4}
\rput(5,0){5}
\rput(6,0){6}
\rput(7,0){7}
\rput(8,0){8}
\rput(9,0){9}
\rput(10,0){10}
\rput(5.5,.5){$j$}
\rput(0,-10){10}
\rput(0,-9){9}
\rput(0,-8){8}
\rput(0,-7){7}
\rput(0,-6){6}
\rput(0,-5){5}
\rput(0,-4){4}
\rput(0,-3){3}
\rput(0,-2){2}
\rput(0,-1){1}
\rput(-.5,-5.5){$i$}
\rput(1,-1){$\circ$}
\rput(2,-1){$\circ$}
\rput(3,-1){$\circ$}
\rput(4,-1){$\circ$}
\rput(5,-1){$\circ$}
\rput(6,-1){$\circ$}
\rput(7,-1){$\circ$}
\rput(8,-1){$\circ$}
\rput(9,-1){$\circ$}
\rput(10,-1){$\circ$}
\rput(2,-2){$\bullet$}
\rput(3,-2){$\bullet$}
\rput(4,-2){$\bullet$}
\rput(5,-2){$\bullet$}
\rput(6,-2){$\circ$}
\rput(7,-2){$\circ$}
\rput(8,-2){$\circ$}
\rput(9,-2){$\circ$}
\rput(10,-2){$\circ$}
\rput(3,-3){$\bullet$}
\rput(4,-3){$\bullet$}
\rput(5,-3){$\bullet$}
\rput(6,-3){$\circ$}
\rput(7,-3){$\circ$}
\rput(8,-3){$\circ$}
\rput(9,-3){$\circ$}
\rput(10,-3){$\circ$}
\rput(4,-4){$\bullet$}
\rput(5,-4){$\bullet$}
\rput(6,-4){$\circ$}
\rput(7,-4){$\circ$}
\rput(8,-4){$\circ$}
\rput(9,-4){$\circ$}
\rput(10,-4){$\circ$}
\rput(5,-5){$\bullet$}
\rput(6,-5){$\circ$}
\rput(7,-5){$\circ$}
\rput(8,-5){$\circ$}
\rput(9,-5){$\circ$}
\rput(10,-5){$\circ$}
\rput(6,-6){$\circ$}
\rput(7,-6){$\circ$}
\rput(8,-6){$\circ$}
\rput(9,-6){$\circ$}
\rput(10,-6){$\circ$}
\rput(7,-7){$\circ$}
\rput(8,-7){$\circ$}
\rput(9,-7){$\circ$}
\rput(10,-7){$\circ$}
\rput(8,-8){$\bullet$}
\rput(9,-8){$\bullet$}
\rput(10,-8){$\bullet$}
\rput(9,-9){$\bullet$}
\rput(10,-9){$\bullet$}
\rput(10,-10){$\bullet$}
\endpspicture
\end{center}
\caption{\label{MunozEaton}A typical configuration of the
  Mu\~noz--Eaton model. An empty (resp.\ filled) circle at row $i$ and
  column $j$ represents the variable $x_{i,j}$ taking value 0 (resp.\
  1).}
\end{figure}

In order to derive the factorization of the probability
\cite{PelJSTAT}, we need first to exploit the locality of
interactions, which allows us to write
\begin{equation}
p(\{x_{i,j}\}) = \frac{p^{(1,2)} p^{(2,3)} \cdots p^{(L-1,L)}}{p^{(2)}
  \cdots p^{(L-1)} },
\label{ME-TMfactoring}
\end{equation}
where $p^{(j)}$ denotes the probability of the $j$th row in
\Fref{MunozEaton} and $p^{(j,j+1)}$ denotes the joint probability of
rows $j$ and $j+1$.

As a second step, consider the effect of the constraints. This is best
understood looking at the following example:
\begin{eqnarray}
p^{(j)}(0, \cdots 0_i, 1_{i+1}, \cdots 1) &=& p^{(j)}_{i,i+1}(0,1) =
  \nonumber \\
&=& \frac{p^{(j)}_{1,2}(0,0) \cdots p^{(j)}_{i,i+1}(0,1) \cdots
  p^{(j)}_{j-1,j}(1,1)}{p^{(j)}_2(0) \cdots p^{(j)}_i(0) 
  p^{(j)}_{i+1}(1) \cdots p^{(j)}_{j-1}(1)}.
\end{eqnarray}
The CVM--like factorization is possible since every factor in the
numerator, except $p^{(j)}_{i,i+1}(0,1)$, cancels with a factor in the
denominator. A similar result can be obtained for the joint
probability of two adjacent rows, and substituting into
\eref{ME-TMfactoring} one eventually gets 
\begin{equation}
p(\{x_{i,j}\}) = \prod_{\alpha \in R} p_\alpha(x_\alpha)^{a_\alpha},
\end{equation}
where $R = \{$square plaquettes, corners (on the diagonal), and their
subclusters$\}$ and $a_\alpha$ is the CVM M\"obius number for cluster
$\alpha$.

\section{Cluster Variation Method as an approximation}
\label{Approx}

In most applications the CVM does not yield exact results, and
hence it is worth investigating its properties as an
approximation. 

An important issue is the choice of maximal clusters, and in
particular the existence of sequence of approximations (that is,
sequence of choices of maximal clusters) with some property of
convergence to the exact results. This has been long studied in the
literature regarding applications to lattice, translation invariant,
systems and will be the subject of subsection \ref{Asymptotic}. In
particular, rigorous results concerning sequences of approximations
which converge to the exact solution have been derived by Schlijper
\cite{Sch83,Sch84,Sch85}, providing a sound theoretical basis for the
earlier investigations by Kikuchi and Brush \cite{KikBru}.

Another important issue is related to the practical determination of
the minima of the CVM variational free energy. In the variational
formulation of statistical mechanics the free energy is convex, but
this property here is lost due to the presence of negative $a_\alpha$
coefficients in the entropy expansion. More precisely, it has been
shown \cite{PakAna} that the CVM variational free energy is convex if
\begin{equation}
\forall S \subseteq R \qquad \sum_{\alpha \in R_S} a_\alpha \ge 0
\qquad R_S = \{ \alpha \in R | \exists \beta \subseteq \alpha, 
\beta \in S \}.
\end{equation}
Similar conditions have been obtained by McEliece and Yildirim
\cite{McEYil} and Heskes, Albers and Kappen \cite{HAK}.

If this is not the case multiple minima can appear, and their
determination can be nontrivial. Several algorithms have been
developed to deal with this problem, falling mainly in two classes:
message--passing algorithms, which will be discussed in subsection
\ref{MessPassAlg} and variational, provably convergent algorithms,
which will be discussed in subsection \ref{VarAlg}.

\subsection{Asymptotic behaviour}
\label{Asymptotic}

The first studies on the asymptotic behaviour of sequences of CVM
approximations are due to Schlijper \cite{Sch83,Sch84,Sch85}. He
showed that it is possible to build sequences of CVM approximations
(that is, sequences of sets of maximal clusters) such that the
corresponding sequence of free energies converge to the exact one, for
a translation--invariant model in the thermodynamic limit. The most
interesting result, related to the transfer matrix idea, is that for a
$d$--dimensional model the maximal clusters considered have to
increase in $d-1$ dimensions only. 

\begin{figure}[h]
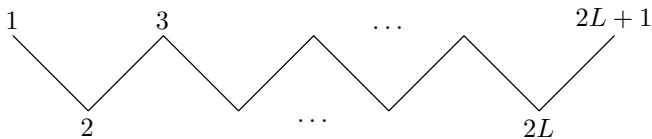

\centertexdraw{
\drawdim cm \linewd 0.02 
\arrowheadsize l:0.3 w:0.15
\arrowheadtype t:V
\move(-2 -2) 
\rlvec(1 -1) \rlvec(1 1)
\rlvec(1 -1) \rlvec(1 1)
\rlvec(1 -1) \rlvec(1 1)
\rlvec(1 -1) \rlvec(1 1)
\textref h:C v:B
\htext(-2 -1.9) {$1$} 
\htext(0 -1.9) {$3$} 
\htext(3 -1.9) {$\ldots$} 
\htext(6 -1.9) {$2L+1$}
\textref h:C v:T 
\htext(-1 -3.1) {$2$} 
\htext(2 -3.1) {$\ldots$} 
\htext(5 -3.1) {$2L$} 
}
\caption{Maximal cluster for the $B_{2L}$ approximation}
\label{KikBruFig}
\end{figure}

These results provided a theoretical justification for the series of
approximations developed by Kikuchi and Brush \cite{KikBru}, who
introduced the $B_{2L}$ series of approximations for
translation--invariant models on the two--dimensional square lattice,
based on zig--zag maximal clusters, as shown in \Fref{KikBruFig}, and
applied it to the zero field Ising model. Based solely on the results
from the $B_2$ (which is equivalent to the plaquette approximation)
and $B_4$ approximations, Kikuchi and Brush postulated a linear
behaviour for the estimate of the critical temperature as a function
of $(2L+1)^{-1}$.

\begin{figure}
\begin{center}
\includegraphics*[scale=.5]{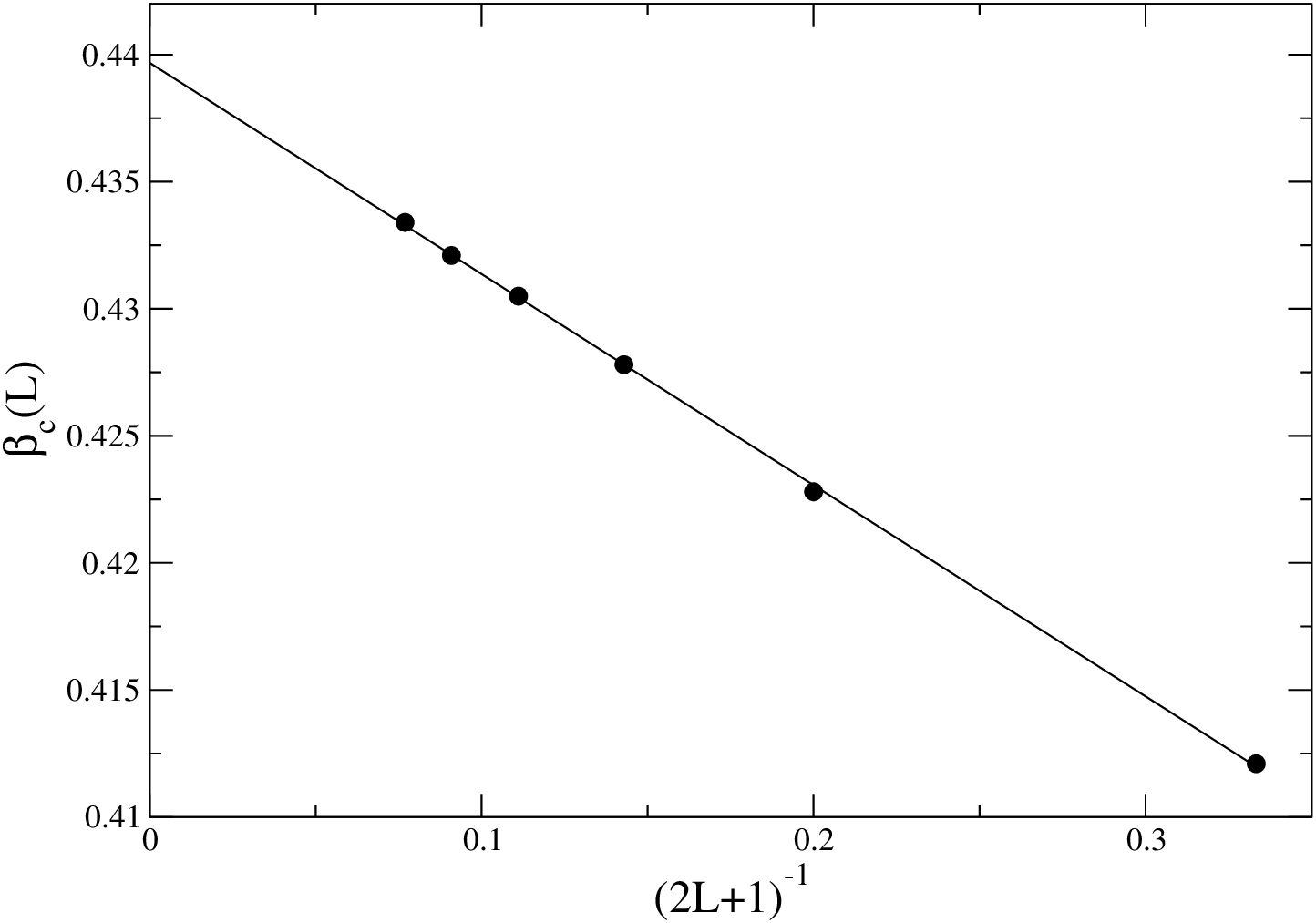}
\end{center}
\caption{\label{B2L-Tc}Inverse critical temperature of the $B_{2L}$
  approximation series}
\end{figure}

In \Fref{B2L-Tc} we have reported the inverse critical temperature as
a function of $(2L+1)^{-1}$ for $L = 1$ to 6. The extrapolated inverse
critical temperature is $\beta_c \simeq 0.4397$, to be compared with
the exactly known $\beta_c = \frac{1}{2} \ln(1 + \sqrt{2}) \simeq
0.4407$.

\begin{figure}
\begin{center}
\includegraphics*[trim = 0 0 0 50, scale=.5]{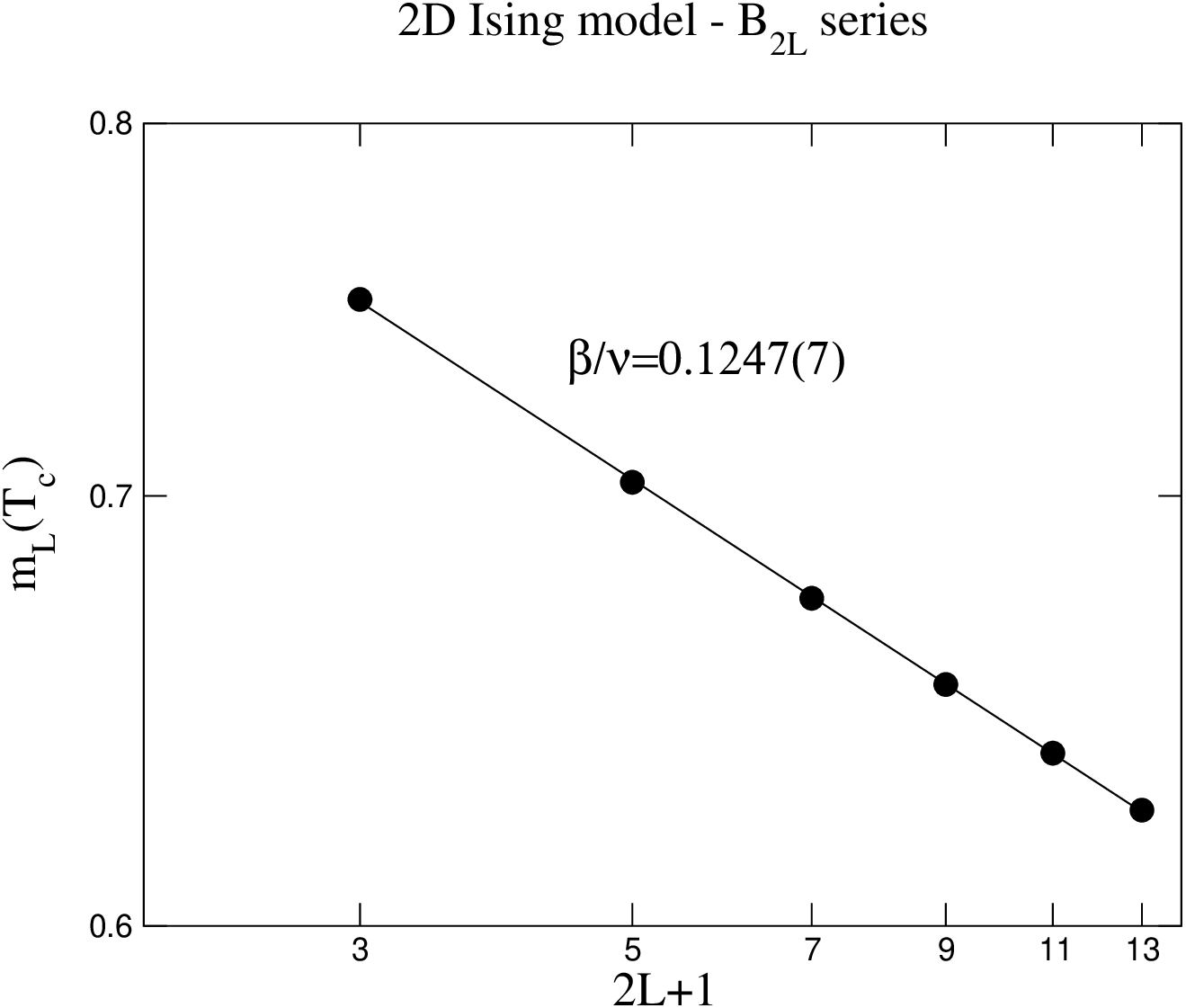}
\end{center}
\caption{\label{FracBetaNu}Critical temperature of the $B_{2L}$
  approximation series}
\end{figure}

It is not our purpose here to make a complete finite size scaling
analysis, in the spirit of the coherent anomaly method (see below), of
the CVM approximation series. We limit ourselves to show the finite
size behaviour of the critical magnetization. More precisely, we have
computed the magnetization of the zero field Ising model on the square
lattice at the exactly known inverse critical temperature, again for
$L = 1$ to 6. \Fref{FracBetaNu} shows that the critical magnetization
vanishes as $(2L+1)^{\beta/\nu}$, and the fit gives a very good
estimate for the exponent, consistent with the exact result $\beta/\nu
= 1/8$.

\begin{figure}
\begin{center}
\includegraphics*[scale=.5]{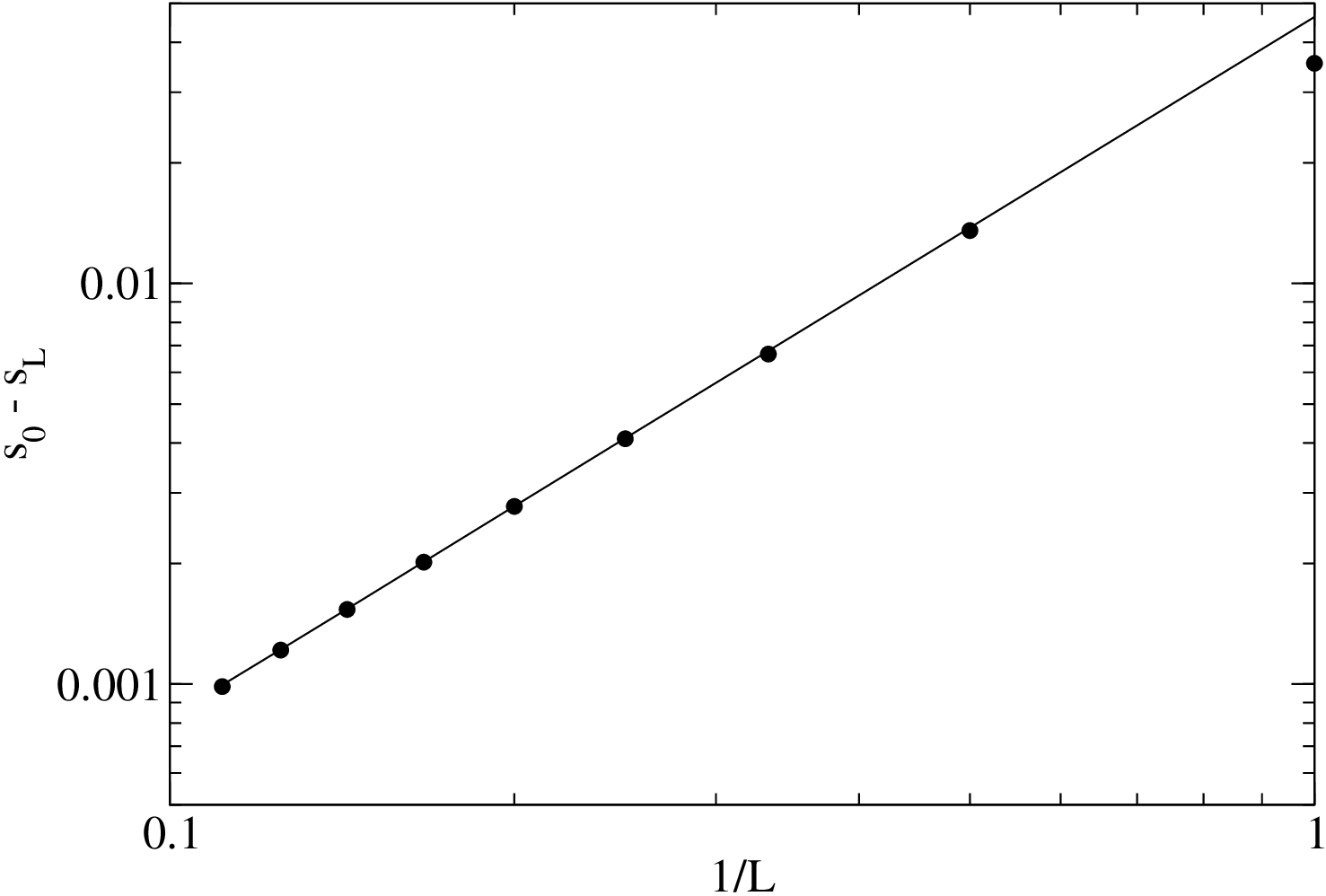}
\end{center}
\caption{\label{TrAFEntropy}Zero temeperature entropy of the triangular
  Ising antiferromagnet in the $B_{2L}$ approximation series}
\end{figure}

As a further illustration of the asymptotic properties of the $B_{2L}$
series we report in \Fref{TrAFEntropy} the zero temperature entropy
(actually the difference between the extrapolated entropy density and
the $B_{2L}$ estimate) of the Ising triangular antiferromagnet as a
function of $1/L$ \cite{PelPre}. It is clearly seen that
asymptotically $s_L = s_0 - a L^{-\psi}$, and the fit yields the
numerical results $s_0 \approx 0.323126$ (the exact value being
$s \approx 0.323066$) and $\psi \approx 1.7512$ (remarkably close to
$7/4$).

An attempt to extract non--classical critical behaviour from high
precision low and high temperature results from CVM was made by the
present author \cite{CVPAM1,CVPAM2,CVPAM3,CVPAM4}, using Pad\'e and
Adler approximants. This work has led to the development of an 18 ($3
\times 3 \times 2$) site cluster approximation for the simple cubic
lattice Ising model \cite{CVPAM4}, which is probably the largest
cluster ever considered. The results obtained for the Ising model are
still compatible with the most recent estimates \cite{PelVic},
although of lower precision.

It has also been considered the possibility of extracting
non--classical critical behaviour from CVM results by means of the
coherent anomaly method, which applies finite size scaling ideas to
series of mean--field--like approximations. A review of these results
can be found in \cite{CAM}.

\subsection{Message--passing algorithms}
\label{MessPassAlg}

In order to describe this class of algorithms it is useful to start
with the Bethe--Peierls approximation (pair approximation of the CVM)
free energy for the Ising model \Eref{Ising}:
\begin{eqnarray}
\fl {\cal F} = - \sum_i h_i \sum_{s_i} s_i p_i(s_i)
- \sum_{\langle i j \rangle} J_{ij} 
\sum_{s_i,s_j} s_i s_j p_{ij}(s_i,s_j) + \nonumber \\ 
\lo + \sum_{\langle i j \rangle} \sum_{s_i,s_j} p_{ij}(s_i,s_j) \ln
p_{ij}(s_i,s_j) 
- \sum_i (d_i-1) \sum_{s_i} p_i(s_i) \ln p_i(s_i) \nonumber \\
\lo + \sum_i \lambda_i \left( \sum_{s_i} p_i(s_i) - 1 \right) 
+ \sum_{\langle i j \rangle} \lambda_{ij} \left( \sum_{s_i,s_j}
p_{ij}(s_i,s_j) - 1 \right)  
+ \nonumber \\
\lo + \sum_{\langle i j \rangle} \left[ \nu_{i,j} \left(
\sum_{s_i} s_i p_i(s_i) - \sum_{s_i,s_j} s_i p_{ij}(s_i,s_j) \right) +
\right. \nonumber \\
\lo + \left.  \nu_{j,i} \left(
\sum_{s_j} s_j p_j(s_j) - \sum_{s_i,s_j} s_j p_{ij}(s_i,s_j) \right)
\right]. 
\label{BetheIsing}
\end{eqnarray}
One can easily recognize the energy terms, the pair entropy, the site
entropy (with a M\"obius number $-(d_i-1)$, where $d_i$ is the degree
of node $i$), and the Lagrange terms corresponding to the
normalization and pair--site compatibility constraints. Observe that,
due the presence of normalization constraints, it is enough to impose
the consistency between the spin expectation values given by the site
and pair probabilities. 

A physical way of deriving message--passing algorithms for the
determination of the stationary points of the above
free energy is through the introduction of the effective field
representation. Consider the interaction $J_{ik}$ and assume that,
whenever this is not taken into account exactly, its effect on
variable $s_i$ can be replaced by an effective field $h_{i,k}$. This
can be made rigorous by observing that the stationarity conditions
\begin{eqnarray}
\frac{\partial {\cal F}}{\partial p_i(s_i)} = 0 \nonumber \\
\frac{\partial {\cal F}}{\partial p_{ij}(s_i,s_j)} = 0
\label{Stationarity}
\end{eqnarray}
can be solved by writing the probabilities as
\begin{eqnarray}
\fl p_i(s_i) &=& \exp\left[ F_i + \left( h_i +
\sum_{k \, {\rm NN} \, i} h_{i,k} \right) s_i \right]
\\
\fl p_{ij}(s_i,s_j) &=& \exp\left[ F_{ij} +
\left( h_i + \sum_{k \, {\rm NN} \, i}^{k \ne j} h_{i,k} \right) s_i  
+ \left( h_j + \sum_{k \, {\rm NN} \, j}^{k \ne i} h_{j,k} \right) s_j +
J_{ij} s_i s_j \right], 
\label{p-vs-heff}
\end{eqnarray}
where the effective fields, and the site and pair free energies $F_i$
and $F_{ij}$, are related to the Lagrange multipliers through
\begin{eqnarray}
\lambda_i &=& (d_i - 1)(1 + F_i) \nonumber \\
\lambda_{ij} &=& - 1 - F_{ij} \nonumber \\
\nu_{i,j} &=& h_i + \sum_{k \, {\rm NN} \, i}^{k \ne j} h_{i,k}.
\end{eqnarray}
$F_i$ and $F_{ij}$ are determined by the normalization, but first of
all the effective fields must be computed by imposing the
corresponding compatibility constraints, which can be cast into the
form
\begin{equation}
h_{i,j} = {\rm tanh}^{-1} \left [
{\rm tanh}\left(h_j +
\sum_{k \, {\rm NN} \, j}^{k \ne i} h_{j,k}\right)
{\rm tanh} J_{ij} \right].
\label{heff-iter}
\end{equation}

This is a set of coupled nonlinear equations which is often solved by
iteration, that is an initial guess is made for the $h_{i,j}$'s,
plugged into the r.h.s.\ of \Eref{heff-iter} which returns a new
estimate, and the procedure is then repeated until a fixed point is
(hopefully) reached. The values of the effective fields at the fixed
point can then be used to compute the probabilities according to
\Eref{p-vs-heff}.

The above equations, and their generalizations at the CVM level, have
been intensively used in the 80's for studying the average behaviour
of models with quenched random interactions, like Ising spin glass
models. This work was started by a paper by Morita \cite{Mor79}, where
an integral equation for the probability distribution of the effective
field, given the probability distributions of the interactions and
fields, was derived. In the general case this integral equation takes
the form
\begin{eqnarray}
\fl p_{i,j}(h_{i,j}) = \int \delta \left( h_{i,j} -
{\rm tanh}^{-1} \left [
{\rm tanh}\left(h_j +
\sum_{k \, {\rm NN} \, j}^{k \ne i} h_{j,k}\right)
{\rm tanh} J_{ij} \right] \right) \times \nonumber \\
\lo \times P_j(h_j) dh_j P_{ij}(J_{ij}) dJ_{ij}
\prod_{k \, {\rm NN} \, j}^{k \ne i} p_{j,k} (h_{j,k}) dh_{j,k},
\label{IntegralEquation}
\end{eqnarray}
with simplifications occurring if the probability distributions can be
assumed to be site--independent, which is the most studied
case. Typical calculations concerned: the determination of elements of
the phase diagrams of Ising spin glass models, through the calculation
of the instability loci of the paramagnetic solution; results in the
zero temperature limit, where solutions with a discrete support are
found; iterative numerical solutions of the integral equation. A
review of this line of research until 1986 can be found in
\cite{Kat86}. It is important to notice that the results obtained by
this approach are equivalent to those by the replica method, at the
replica symmetric level.

The effective field approach is reminiscent of the message--passing
procedure at the heart of the belief propagation (BP) algorithm, and
indeed the messages 
appearing in this algorithm are related, in the Ising case, to the
effective fields by $m_{\langle i j \rangle \to i}(s_i) =
\exp(h_{i,j} s_i)$, where $m_{\langle i j \rangle \to i}(s_i)$ denotes
a message going from the NN pair $\langle i j \rangle$ to node $i$.

In order to derive the BP algorithm consider the Bethe--Peierls
approximation for a model with variable nodes $i$ and factor nodes
$a$. The variables $s_i$ need not be limited to two states
and the Hamiltonian is written in the general form \Eref{HsumHa}.

The CVM free energy, with the normalization and compatibility
constraints, can then be written as
\begin{eqnarray}
\fl {\cal F} = - \sum_a \sum_{\bi{s_a}} H_a(\bi{s_a}) p_a(\bi{s_a})
+ \nonumber \\ \lo 
+ \sum_a \sum_{\bi{s_a}} p_a(\bi{s_a}) \ln p_a(\bi{s_a})
- \sum_i (d_i-1) \sum_{s_i} p_i(s_i) \ln p_i(s_i) + \nonumber \\
\lo + \sum_i \lambda_i \left( \sum_{s_i} p_i(s_i) - 1 \right) 
+ \sum_a \lambda_a \left( 
\sum_a \sum_{\bi{s_a}} p_a(\bi{s_a}) - 1 \right) 
+ \nonumber \\
\lo + \sum_a \sum_{i \in a}
\sum_{s_i} \mu_{a,i}(s_i) \left( p_i(s_i) - \sum_{\bi{s_{a \setminus i}}}
p_a(\bi{s_a}) \right),
\label{BetheFree}
\end{eqnarray}
where $\bi{s_{a \setminus i}}$ denotes the set of variables entering
factor node $a$, except $i$.

The stationarity conditions 
\begin{eqnarray}
\frac{\partial {\cal F}}{\partial p_i(s_i)} = 0 \nonumber \\
\frac{\partial {\cal F}}{\partial p_a(\bi{s_a})} = 0
\end{eqnarray}
can then be solved, in
analogy with \Eref{p-vs-heff}, by
\begin{eqnarray}
p_i(s_i) &=& \frac{1}{Z_i} \prod_{i \in a}
m_{a \to i}(s_i) \nonumber \\
p_a(\bi{s_a}) &=& \frac{1}{Z_a} \psi_a(\bi{s_a}) 
\prod_{k \in a} \prod_{k \in b}^{b \ne a} m_{b \to k}(s_k).
\label{p-vs-mess}
\end{eqnarray}
In the particular case of an Ising model with pairwise interactions,
the previously mentioned relationship between messages and effective
fields is evident from the above equation.

Now $Z_i$ and $Z_a$ take care of normalization, and the messages
$m_{a \to i}(s_i)$ are related to the Lagrange multipliers by
\begin{equation}
\mu_{a,k}(s_k) = \sum_{k \in b}^{b \ne a} \ln m_{b \to k}(s_k).
\end{equation}

Notice that the messages can be regarded as exponentials of a new set
of Lagrange multipliers, with the constraints rewritten as in the
following free energy
\begin{eqnarray}
\fl {\cal F} = - \sum_a \sum_{\bi{s_a}} H_a(\bi{s_a}) p_a(\bi{s_a})
+ \nonumber \\ \lo 
+ \sum_a \sum_{\bi{s_a}} p_a(\bi{s_a}) \ln p_a(\bi{s_a})
- \sum_i (d_i-1) \sum_{s_i} p_i(s_i) \ln p_i(s_i) + \nonumber \\
\lo + \sum_i \lambda_i \left( \sum_{s_i} p_i(s_i) - 1 \right) 
+ \sum_a \lambda_a \left( 
\sum_a \sum_{\bi{s_a}} p_a(\bi{s_a}) - 1 \right) 
+ \nonumber \\
\lo + \sum_a \sum_{i \in a} 
\sum_{s_i} \ln m_{a \to i}(s_i) \left( (d_i - 1) p_i(s_i) - 
\sum_{i \in b}^{b \ne a} \sum_{\bi{s_{b \setminus i}}}
p_b(\bi{s_b}) \right).
\label{BetheFreeRot}
\end{eqnarray}

Again, imposing compatibility between variable nodes and factor nodes,
one gets a set of coupled equations for the messages which, leaving
apart normalization, read
\begin{equation}
m_{a \to i}(s_i) \propto \sum_{\bi{s_{a \setminus i}}} \psi_a(\bi{s_a}) 
\prod_{k \in a}^{k \ne i} \prod_{k \in b}^{b \ne a} m_{b \to k}(s_k).
\label{BP-mess-upd}
\end{equation}

The above equations, and their iterative solution, are the core of the
BP algorithm. Also, their structure justifies the name ``Sum-Product''
\cite{Kschischang}, which is often given them in the literature on
probabilistic graphical models, and the corresponding term
``Max-Product'' for their zero temperature limit.

There are several issues which must be considered when discussing the
property of an iterative algorithm based on \Eref{BP-mess-upd}. First
of all, one could ask whether messages have to be updated sequentially
or in parallel. This degree of freedom does not affect the fixed
points of the algorithm, but it affects the dynamics. This issue has
been considered in some depth by Kfir and Kanter \cite{KfirKanter} in
the context of the decoding of error--correcting codes. In that case
they showed that the sequential update results in twice faster
convergence with respect to the parallel update.

Convergence is however not guaranteed if the underlying graph is not
tree--like, that is if the pair approximation of the CVM is not
exact. This issue has been investigated theoretically by Tatikonda and
Jordan \cite{TatiJor}, Mooij and Kappen \cite{MooKap}, Ihler et al
\cite{Ihler}, who derived sufficient conditions for convergence, and
by Heskes \cite{Heskes2004}, who derived sufficient conditions for the
uniqueness of the fixed point. In practice it is typically observed
that the BP algorithm converges if the frustration due to competitive
interactions, like those characteristic of spin--glass or constraint
satisfaction models, is not too large. In some cases, the trick of
damping, or inertia, can help extending the convergence domain. The
trick consists in taking the updated message equal to a weighted
(possibly geometrical) average of the old message and the new one
given by \Eref{BP-mess-upd}. The convergence domain of the BP
algorithm has been determined for several problems, like
satisfiability \cite{SPSAT}, graph colouring \cite{SPCOL}, error
correcting codes \cite{KabSaaLDPCC} and spin glasses
\cite{SG-BP-conv}. Within its convergence domain, the BP algorithm is
indeed very fast, and this is its real strength. See the next
subsection for some performance tests and a comparison with provably
convergent algorithms.

Once a fixed point has been obtained it is worth asking whether this
corresponds to a minimum of the free energy or not.  This has been
partially solved by Heskes \cite{Heskes}, who has shown that stable
fixed points of the belief propagation algorithm are minima of the CVM
pair approximation free energy, but the converse is not necessarily
true. Actually, examples can be found of minima of the free energy
which correspond to unstable fixed points of the belief propagation
algorithm.

An important advancement in this topic is the {\em generalized belief
propagation} (GBP) algorithm by Yedidia and coworkers
\cite{Yed01}. The fixed points of the GBP algorithm for a certain
choice of clusters correspond to stationary points of the CVM free
energy at the approximation level corresponding by the same choice of
clusters or, more generally, of a region graph free energy. Actually,
for a given choice of clusters, different GBP algorithms can be
devised. Here only the so--called {\em parent to child} GBP algorithm
\cite{Yed04} will be considered. Other choices are described in
\cite{Yed04}. 

In order to better understand this algorithm, notice a few
characteristics of the belief propagation algorithm. First of all,
looking at the probabilities \Eref{p-vs-mess} one can say that a
variable node receives messages from all the factor nodes it belongs
to, while a factor node $a$ receives messages from all the other
factor nodes to which its variable nodes $i \in a$ belong. In
addition, the constraint corresponding to the message $m_{a \to
i}(s_i)$ (see \Eref{BetheFreeRot}) can be written as
\begin{equation}
\sum_{\bi{s_{a \setminus i}}} p_a(\bi{s_a}) = 
\sum_{i \in b} \sum_{\bi{s_{b \setminus i}}} p_b(\bi{s_b})
- (d_i - 1) p_i(s_i).
\end{equation}

The parent to child GBP algorithm generalizes these characteristics in
a rather straightforward way. First of all, messages $m_{\alpha \to
\beta}(\bi{s_\beta})$ ($\beta \subset \alpha$) are introduced from
regions (parent regions) to subregions (child regions). Then, the
probability of a region takes into account messages coming from outer
regions to itself and its subregions. Finally, exploiting the property
\Eref{MobiusNumbers} of the M\"obius numbers, the constraint
corresponding to $m_{\alpha \to \beta}(\bi{s_\beta})$ is written in the
form
\begin{equation}
\sum_{\alpha \subseteq \gamma \in R} a_\gamma \sum_{\bi{s_{\gamma \setminus
\beta}}} p_\gamma(\bi{s_\gamma}) = \sum_{\beta \subseteq \gamma \in R}
a_\gamma \sum_{\bi{s_{\gamma \setminus \beta}}}
p_\gamma(\bi{s_\gamma}).
\end{equation}
It can be shown \cite{Yed04} that this new set of constraints is
equivalent to the original one. 

To make this more rigorous, consider the free energy given by
Equations (\ref{CVMFree}) and (\ref{ClusterFree}), with the above
compatibility constraints (with Lagrange multipliers $\ln m_{\alpha
\to \beta}(\bi{s_\beta})$) and the usual normalization constraints
(with multipliers $\lambda_\alpha$). 
One obtains
\begin{eqnarray}
\fl {\cal F} = \sum_{\gamma \in R} a_\gamma \sum_{\bi{s_\gamma}} \left[
  p_\gamma(\bi{s_\gamma}) H_\gamma(\bi{s_\gamma}) + p_\gamma(\bi{s_\gamma}) \ln
  p_\gamma(\bi{s_\gamma}) \right] + \sum_{\gamma \in R} \lambda_\gamma
  \left[ \sum_{\bi{s_\gamma}} p_\gamma(\bi{s_\gamma}) - 1 \right] +
  \nonumber \\ 
\fl + \sum_{\beta \subset \alpha \in R} \sum_{\bi{s_\beta}} 
\ln m_{\alpha \to \beta}(\bi{s_\beta}) \left[
\sum_{\alpha \subseteq \gamma \in R} a_\gamma \sum_{\bi{s_{\gamma \setminus
\beta}}} p_\gamma(\bi{s_\gamma}) - \sum_{\beta \subseteq \gamma \in R}
a_\gamma \sum_{\bi{s_{\gamma \setminus \beta}}}
  p_\gamma(\bi{s_\gamma}) \right],
\end{eqnarray}
where it is not necessary to put all the possible $\alpha \to \beta$
compatibility constraints, but it is enough to put those which satisfy
$a_\alpha \ne 0$, $a_\beta \ne 0$ and $\beta$ is a direct subregion of
$\alpha$, that is there is no region $\gamma$ with $a_\gamma \ne 0$
such that $\beta \subset \gamma \subset \alpha$. Notice also that the
Lagrange term corresponding to the $\alpha \to \beta$ constraint can
be written as
\begin{equation}
- \ln m_{\alpha \to \beta}(\bi{s_\beta}) \sum_{\beta \subseteq \gamma
  \in R}^{\alpha \nsubseteq \gamma} a_\gamma \sum_{\bi{s_{\gamma
      \setminus \beta}}} p_\gamma(\bi{s_\gamma}).
\end{equation}

The stationarity conditions
\begin{equation}
\frac{\partial {\cal F}}{\partial p_\gamma(\bi{s_\gamma})} = 0
\end{equation}
can then be solved, leaving apart normalization, by
\begin{equation}
p_\gamma(\bi{s_\gamma}) \propto \exp\left[ - H_\gamma(\bi{s_\gamma})
  \right] \prod_{\beta \subseteq \gamma} \prod_{\beta \subset \alpha
  \in R}^{\alpha \nsubseteq \gamma} m_{\alpha \to
  \beta}(\bi{s_\beta}),
\label{GBP-p-vs-mess}
\end{equation}
where $\bi{s_\beta}$  denotes the restriction of $\bi{s_\gamma}$  to
subregion $\beta$. 

Finally, message update rules can be derived again by the compatibility
constraints, though some care is needed, since in the general case
these constraints are not immediately solved with respect to the
(updated) messages, as it occurs in the derivation of
\Eref{BP-mess-upd}. Here one obtains a coupled set of equations in the
updated messages, which can be solved starting from the constraints
involving the smallest clusters. 

\begin{figure}
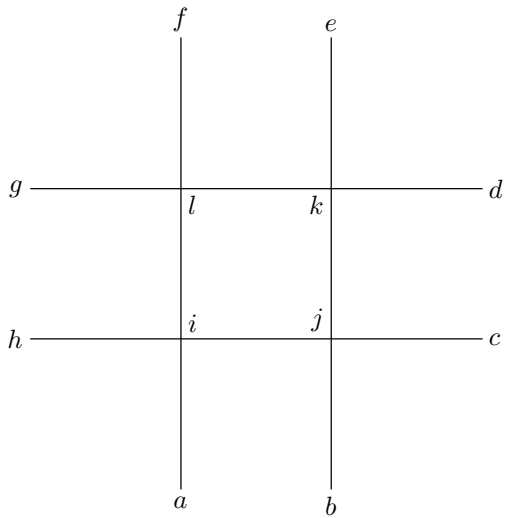

\centertexdraw{
\drawdim cm \linewd 0.02 
\arrowheadsize l:0.3 w:0.15
\arrowheadtype t:V
\move(0 2) \lvec(6 2) 
\move(0 4) \lvec(6 4)
\move(2 0) \lvec(2 6) 
\move(4 0) \lvec(4 6) 
\textref h:L v:B \htext(2.1 2.1) {$i$}
\textref h:R v:B \htext(3.9 2.1) {$j$}
\textref h:L v:T \htext(2.1 3.9) {$l$}
\textref h:R v:T \htext(3.9 3.9) {$k$}
\textref h:C v:T \htext(2 -0.1) {$a$}
\textref h:C v:T \htext(4 -0.1) {$b$}
\textref h:L v:C \htext(6.1 2) {$c$}
\textref h:L v:C \htext(6.1 4) {$d$}
\textref h:C v:B \htext(4 6.1) {$e$}
\textref h:C v:B \htext(2 6.1) {$f$}
\textref h:R v:C \htext(-0.1 4) {$g$}
\textref h:R v:C \htext(-0.1 2) {$h$}
}
\caption{\label{SquareLatticePortion}A portion of the square lattice}
\end{figure}

An example can be helpful here. Consider a model defined on a regular
square lattice, with periodic boundary conditions, and the CVM square
approximation, that is the approximation obtained by taking the
elementary square plaquettes as maximal clusters. The entropy
expansion contains only terms for square plaquettes (with M\"obius
numbers 1), NN pairs (M\"obius numbers -1) and single nodes (M\"obius
numbers 1), as in \Eref{SquareEntropy}. A minimal set of compatibility
constraints includes node--pair and pair--square constraints, and one
has therefore to deal with square--to--pair and pair--to--node
messages, which will be denoted by $m_{ij,kl}(s_i,s_j)$ and
$m_{i,j}(s_i)$ respectively. With reference to the portion of the
lattice depicted in \Fref{SquareLatticePortion} the probabilities,
according to \Eref{GBP-p-vs-mess}, can be written as
\begin{eqnarray}
\fl p_i(s_i) \propto \exp[-H_i(s_i)] \, m_{i,a}(s_i) \, m_{i,j}(s_i)
\, m_{i,l}(s_i) \, m_{i,h}(s_i), \nonumber \\
\fl p_{ij}(s_i,s_j) \propto \exp[-H_{ij}(s_i,s_j)] \, m_{i,a}(s_i) \, 
m_{i,l}(s_i) \, m_{i,h}(s_i) \times \nonumber \\
\lo \times m_{j,b}(s_j) \, m_{j,c}(s_j) \, m_{j,k}(s_j) \,
m_{ij,ab}(s_i,s_j) \,
m_{ij,lk}(s_i,s_j), \nonumber \\
\fl p_{ijkl}(s_i,s_j,s_k,s_l) \propto \exp[-H_{ijkl}(s_i,s_j,s_k,s_l)]
\, m_{i,a}(s_i) \, m_{i,h}(s_i) \, m_{j,b}(s_j) \, m_{j,c}(s_j) \times
\nonumber \\
\lo \times m_{k,d}(s_k) \, m_{k,e}(s_k) \, m_{l,f}(s_l) \,
m_{l,g}(s_l) \times \nonumber \\
\lo \times m_{ij,ab}(s_i,s_j) \, 
m_{jk,cd}(s_j,s_k) \, m_{kl,ef}(s_k,s_l) \, m_{lj,gh}(s_l,s_j).
\end{eqnarray}

Imposing node--pair and pair--square constraints one gets equations
like
\begin{eqnarray}
\fl \exp[-H_i(s_i)] \, m_{i,j}(s_i) 
\propto \sum_{s_j} \exp[-H_{ij}(s_i,s_j)] \times \nonumber \\
\lo \times m_{j,b}(s_j) \, m_{j,c}(s_j) \,
m_{j,k}(s_j) \, m_{ij,ab}(s_i,s_j) \, m_{ij,lk}(s_i,s_j), \nonumber \\
\fl \exp[-H_{ij}(s_i,s_j)] \, m_{i,f}(s_i) \, m_{j,k}(s_j) \,
m_{ij,lk}(s_i,s_j) \propto \sum_{s_k,s_l}
\exp[-H_{ijkl}(s_i,s_j,s_k,s_l)]  \times \nonumber \\
\lo \times m_{k,d}(s_k) \,
m_{k,e}(s_k) \, m_{l,f}(s_l) \, m_{l,g}(s_l) \times \nonumber \\
\lo \times m_{jk,cd}(s_j,s_k) \, m_{kl,ef}(s_k,s_l) \, m_{lj,gh}(s_l,s_j).
\end{eqnarray}
The above equations can be viewed as a set of equations in the updated
messages at iteration $t+1$, appearing in the l.h.s., given the
messages at iteration $t$, appearing in the r.h.s.. It is clear that
one has first to calculate the updated pair--to--site messages
according to the first equation, and then the updated square--to--pair
messages according to the second one, using in the l.h.s.\ the updated
pair--to--site messages just obtained.

GBP (possibly with damping) typically exhibits better convergence
properties (and greater accuracy) than BP, but the empirical rule that
a sufficient amount of frustration can make it not convergent is valid
also for GBP. It is therefore fundamental to look for provably
convergent algorithms, which will be discussed in the next
subsection. A variation of the BP algorithm, the conditioned
probability (CP) algorithm, with improved convergence properties, has
recently been introduced \cite{MP-Prop}. The extension of this
algorithm beyond the BP level is however not straightforward.

We conclude the present subsection by mentioning that 
techniques like the Thouless--Anderson--Palmer equations, or the
cavity method, both widely used in the statistical physics of spin
glasses, are strictly related to the Bethe--Peierls approximation.

The Thouless--Anderson--Palmer \cite{TAP} equations can be derived
from the Bethe--Peierls free energy for the Ising model, through the
so-called Plefka expansion \cite{Plefka}. One has first to write the
free energy as a function of magnetizations and nearest--neighbour
correlations through the parameterization
\begin{equation}
p_i(s_i) = \frac{1 + s_i m_i}{2} \qquad
p_{ij}(s_i,s_j) = \frac{1 + s_i m_i + s_j m_j + s_i s_j c_{ij}}{4},
\end{equation}
then to solve analytically the stationarity conditions with respect to
the $c_{ij}$'s and finally to expand to second order in the inverse
temperature. 

Finally, the cavity method \cite{MezPar86,MezPar87,MezPar01} is
particularly important since it allows to deal with replica symmetry
breaking. The cavity method, though historically derived in a
different way, can be regarded as an alternative choice of messages
and effective fields in the Bethe--Peierls approximation. With
reference to \Eref{p-vs-mess}, introduce messages $m_{k \to a}(s_k)$
from variable nodes to factor nodes according to
\begin{equation}
m_{k \to a}(s_k) = \prod_{k \in b}^{b \ne a} m_{b \to k}(s_k).
\end{equation}
Then the probabilities \Eref{p-vs-mess} become
\begin{eqnarray}
p_i(s_i) &=& \frac{1}{Z_i} \prod_{i \in a}
m_{a \to i}(s_i) \nonumber \\
p_a(\bi{s_a}) &=& \frac{1}{Z_a} \psi_a(\bi{s_a}) 
\prod_{k \in a} m_{k \to a}(s_k),
\end{eqnarray}
and the message update equations (\ref{BP-mess-upd}) become
\begin{equation}
m_{a \to i}(s_i) \propto \sum_{\bi{s_{a \setminus i}}} \psi_a(\bi{s_a}) 
\prod_{k \in a}^{k \ne i} m_{k \to a}(s_k).
\end{equation}
The effective fields corresponding to the factor--to--variable
messages $m_{a \to i}(s_i)$ are usually called cavity biases, while
those corresponding to the variable--to--factor messages $m_{i \to
a}(s_i)$ are called cavity fields. In the Ising example above a factor
node is just a pair of NNs and cavity biases reduce to effective
fields $h_{i,j}$, while cavity fields take the form
$\displaystyle{\sum_{k {\rm NN} i}^{k \ne j} h_{i,k}}$.

The cavity method admits an extension to cases where one step of
replica symmetry breaking occurs \cite{MezPar01,MezPar03}. In such a
case one assumes that there exist many states characterized by
different values of the cavity biases and fields, and introduces the
probability distributions of cavity biases and fields over the
states. From the above message update rules one can then derive
integral equations, similar to \Eref{IntegralEquation}, for the
distributions. These integral equations can in principle be solved by
iterative population dynamics algorithms, but most often one restricts
to the zero temperature case, where these distributions have a
discrete support.

The zero temperature case is particularly relevant for hard
combinatorial optimization problems, where 1--step replica symmetry
breaking corresponds to clustering of solutions. Clustering means that
the space of solutions becomes disconnected, made of subspaces which
cannot be reached from one another by means of local moves, and hence
all local algorithms, like BP or GBP, are bound to fail. The cavity
method has been used to solve these kind of problems in the framework
of the survey propagation algorithm \cite{SPScience}, which has been
shown to be a very powerful tool for constraint satisfaction problems
like satisfiability \cite{SPSAT} and colouring \cite{SPCOL} defined on
finite connectivity random graphs. These graphs are locally tree--like
and therefore all the analysis can be carried out at the
Bethe--Peierls level. A sort of generalized survey propagation capable
of dealing with short loops would really be welcome, but it seems that
realizability issues are crucial here and replica symmetry breaking
can only be introduced when CVM gives an exact solution.

A different approach, still aimed to generalize the BP algorithm to
situations where replica symmetry breaking occurs, has been suggested
by van Mourik \cite{Jort}, and is based on the analysis of the time
evolution of the BP algorithm. 

\subsection{Variational algorithms}
\label{VarAlg}

In the present subsection we discuss algorithms which update
probabilities instead of messages. At every iteration a new estimate
of probabilities, and hence of the free energy, is obtained. These
algorithms are typically provably convergent, and the proof is based
on showing that the free energy decreases at each iteration. This is
of course not possible with BP and GBP algorithms, where the
probabilities and the free energy can be evaluated only at the fixed
point. The price one has to pay is that in variational algorithms one
has to solve the compatibility constraints at every iteration, and
therefore these are double loop algorithms, where the outer loop is
used to update probabilities and the inner loop is used to solve the
constraints.

The natural iteration method (NIM) \cite{Kik74,Kik76} is the oldest
algorithm specifically designed to minimize the CVM variational free
energy. It was originally introduced \cite{Kik74} in the context of
homogeneous models, for the pair and tetrahedron (for the fcc lattice)
approximations. In such cases the compatibility constraints are
trivial. Later \cite{Kik76} it was generalized to cases where the
compatibility constraints cannot be solved trivially. An improved
version of the algorithm, with tunable convergence properties,
appeared in \cite{KiKoKa} and its application is described in some
detail also in \cite{3CVM}, where higher order approximations are
considered.

The algorithm is based on a double loop scheme, where the inner loop
is used to solve the compatibility constraints, so that at each
iteration of the outer loop a set of cluster probabilities which
satisfy the constraints is obtained. 

Proof of convergence, based on showing that the free energy decreases
at every outer loop iteration, exist in many cases, but it has also
been shown that there are non--convergent cases, like the
four--dimensional Ising model \cite{Pretti} in the hypercube
approximation. 

We do not discuss in detail this algorithm since it is rather slow,
and better alternatives have been recently developed. 

A first step in this direction was the {\it concave--convex procedure}
(CCCP) by Yuille \cite{Yuille}, who started from the observation that
the non--convergence problems of message--passing algorithms arise
from concave terms in the variational free energy, that is from the
entropy of clusters with negative M\"obius numbers. His idea was then
to split the CVM free energy into a convex and a concave part,
\begin{equation}
{\cal F}(\{p_\alpha\}) = {\cal F}_{\rm vex}(\{p_\alpha\}) + 
{\cal F}_{\rm cave}(\{p_\alpha\}),
\label{CCCPsplit}
\end{equation}
and to write the update equations to be iterated to a fixed point
as
\begin{equation}
\nabla {\cal F}_{\rm vex}(\{p_\alpha^{(t+1)}\}) = - 
\nabla {\cal F}_{\rm cave}(\{p_\alpha^{(t)}\}),
\label{CCCPiter}
\end{equation}
where $p_\alpha^{(t)}$ and $p_\alpha^{(t+1)}$ are successive
iterates. In order to solve the compatibility constraints, at each
iteration of \Eref{CCCPiter}, the Lagrange multipliers enforcing the
constraints are determined by another iterative algorithm where one
solves for one multiplier at a time, and it can be shown that the free
energy decreases at each outer loop iteration. Therefore we have another double
loop algorithm, which is provably convergent, faster than NIM (as we
shall see below), and allows some freedom in the splitting
between convex and concave parts.

A more general and elegant formalism, which will be described in the
following, has however been put forward by Heskes, Albers and Kappen
(HAK) \cite{HAK}. Their basic idea is to consider a sequence of convex
variational free energies such that the sequence of the corresponding
minima tends to the minimum of the CVM free energy. More precisely, if
the CVM free energy ${\cal F}(\{ p_\alpha, \alpha \in R \})$ is
denoted for simplicity by ${\cal F}(p)$, they consider a function
${\cal F}_{\rm conv}(p,p')$, convex in $p$, with the properties
\begin{eqnarray}
{\cal F}_{\rm conv}(p,p') \ge {\cal F}(p), \nonumber \\
{\cal F}_{\rm conv}(p,p) = {\cal F}(p).
\end{eqnarray}
The algorithm is then defined by the update rule for the probabilities
\begin{equation}
p^{(t+1)} = {\rm arg}\min_{p} {\cal F}_{\rm conv}(p,p^{(t)}),
\label{HAKouter}
\end{equation}
and it is easily proved that the free energy decreases at each
iteration and that a minimum of the CVM free energy is recovered at
the fixed point. 

A lot of freedom is left in the definition of ${\cal F}_{\rm conv}$,
and strategies of varying complexity and speed can be obtained. NIM
(when convergent) and CCCP can also be recovered as special cases. 
The general framework is based on the following three properties. 
\begin{enumerate}
\item If $\beta \subset \alpha$, then
\begin{equation}
- S_\alpha + S_\beta = \sum_{\bi{s_\alpha}} p_\alpha(\bi{s_\alpha}) \ln
p_\alpha(\bi{s_\alpha}) - \sum_{\bi{s_\beta}} p_\beta(\bi{s_\beta})
\ln p_\beta(\bi{s_\beta})
\end{equation}
is convex over the constraint set, i.e.\ it is a convex function of
$p_\alpha$ and $p_\beta$ if these satisfy the compatibility constraint
\Eref{CompConstr}.
\item The linear bound 
\begin{equation}
S_\beta = - \sum_{\bi{s_\beta}} p_\beta(\bi{s_\beta}) \ln
  p_\beta(\bi{s_\beta}) \le - \sum_{\bi{s_\beta}}
  p_\beta(\bi{s_\beta}) \ln p'_\beta(\bi{s_\beta}) = S'_\beta
\end{equation}
holds, with equality only for $p'_\beta = p_\beta$
\item If $\gamma \subset \beta$, and $p_\beta$ and $p_\gamma$
  ($p'_\beta$ and $p'_\gamma$) satisfy the compatibility constraints,
  the bound
\begin{eqnarray}
\fl S_\beta - S_\gamma = - \sum_{\bi{s_\beta}} p_\beta(\bi{s_\beta}) \ln
  p_\beta(\bi{s_\beta}) + \sum_{\bi{s_\gamma}} p_\gamma(\bi{s_\gamma})
  \ln p_\gamma(\bi{s_\gamma}) \le \nonumber \\
\lo \le - \sum_{\bi{s_\beta}}
  p_\beta(\bi{s_\beta}) \ln p'_\beta(\bi{s_\beta}) +
  \sum_{\bi{s_\gamma}} p_\gamma(\bi{s_\gamma}) \ln
  p'_\gamma(\bi{s_\gamma}) = S'_\beta - S'_\gamma
\end{eqnarray}
holds, and it is tighter than the previous bound. A tighter bound
typically entail faster convergence. 
\end{enumerate}

In order to give an example, consider again the CVM square
approximation for a model on a regular square lattice with periodic
boundary conditions and focus on the entropy part of the free energy,
which according to the entropy expansion
\Eref{SquareEntropy} has the form
\begin{equation}
\fl - \sum_{\opensquare} S_{\opensquare} + \sum_{\langle i j \rangle}
S_{ij} - \sum_i S_i = \sum_{\opensquare} p_{\opensquare} \ln
p_{\opensquare} - \sum_{\langle i j \rangle} p_{ij} \ln p_{ij} +
\sum_i p_i \ln p_i.
\end{equation}
This contains both convex (from square and site entropy) and concave
terms (from pair entropy). Notice that the numbers of plaquettes is
the same as the number of sites, while there are two pairs (e.g.\
horizontal and vertical) per site. This implies that the free energy
is not convex over the constraint set.

Several bounding schemes are possible to define ${\cal F}_{\rm
conv}$. For instance, one can obtain a function which is just convex
over the constraint set by applying property (iii) to the site terms
and half the pair terms, with the result
\begin{equation}
\fl - \sum_{\opensquare} S_{\opensquare} + \sum_{\langle i j \rangle}
S_{ij} - \sum_i S_i \le - \sum_{\opensquare} S_{\opensquare} +
\frac{1}{2} \sum_{\langle i j \rangle} S_{ij} +
\frac{1}{2} \sum_{\langle i j \rangle} S'_{ij} - \sum_i S'_i.
\label{JustConvex}
\end{equation}
In the following the HAK algorithm will always be used with this
bounding scheme. 

The NIM can be obtained if, starting from the above expression, one
applies property (ii) to the not yet bounded pair terms, with the
result 
\begin{equation}
\fl - \sum_{\opensquare} S_{\opensquare} + \sum_{\langle i j \rangle}
S_{ij} - \sum_i S_i \le - \sum_{\opensquare} S_{\opensquare} +
\sum_{\langle i j \rangle} S'_{ij} - \sum_i S'_i.
\end{equation}
This is clearly a looser bound than the previous one, and hence it
leads to a (much) slower algorithm. In the general case, the NIM
(which of course was formulated in a different way) can be obtained by
bounding all entropy terms except those corresponding to the maximal
clusters. This choice does not always lead to a convex bound (though
in most practically relevant cases this happens) and hence convergence
is not always guaranteed.

The CCCP recipe corresponds to bounding every convex ($a_\beta < 0$)
term by
\begin{equation}
- a_\beta S_\beta \le - S_\beta + (1 - a_\beta) S'_\beta,
\end{equation}
using property (ii). In the present case this gives
\begin{equation}
\fl - \sum_{\opensquare} S_{\opensquare} + \sum_{\langle i j \rangle}
S_{ij} - \sum_i S_i \le - \sum_{\opensquare} S_{\opensquare} -
\sum_{\langle i j \rangle} S_{ij} +
2 \sum_{\langle i j \rangle} S'_{ij} - \sum_i S_i,
\end{equation}
which is convex independently of the constraints, and hence the bound
is again looser than \Eref{JustConvex}

In all cases one is left with a double loop algorithm, the outer loop
being defined by the update rule for probabilities, and the inner loop
being used for the minimization involved in \Eref{HAKouter}. This
minimization is simpler than the original problem, since the function
to be minimized is convex. In each of the above schemes a particular
technique was proposed for the convex minimization in the inner loop,
and here these will not be covered in detail. 

A point which is important to notice here is that the bounding
operation gives a new free energy which is structurally different from
a CVM free energy. It must be minimized with respect to $p$ at fixed
$p'$ and, viewed as a function of $p$, it contains an entropy
expansion with coefficients $\tilde a_\beta$ which do not satisfy
anymore the M\"obius relation (\ref{MobiusNumbers}) (for instance, in
the ``just convex over the constraint set'' scheme, we have
$a_{\opensquare} = 1$, $a_{ij} = -1/2$ and $a_i = 0$). This means that
a message--passing algorithm like parent--to--child GBP, which relies
on the M\"obius property, cannot be applied. In \cite{HAK} a different
message--passing algorithm, which can still be viewed as a GBP
algorithm, is suggested.

Observe also that there are entropy--like terms $S'_\beta$ which are
actually linear in $p_\beta$ and must therefore be absorbed in the
energy terms.

The main reason for investigating these double loop, provably
convergent algorithms, is the non--convergence of BP and GBP in
frustrated cases. Since BP and GBP, when they converge, are the
fastest algorithms for the determination of the minima of the CVM free
energy, it is worth making some performance tests to evaluate the
speed of the various algorithms. The CPU times reported below refer to
an Intel Pentium 4 processor at 3.06 GHz, using g77 under GNU/Linux.

\begin{figure}
\begin{center}
\includegraphics*[scale=.5]{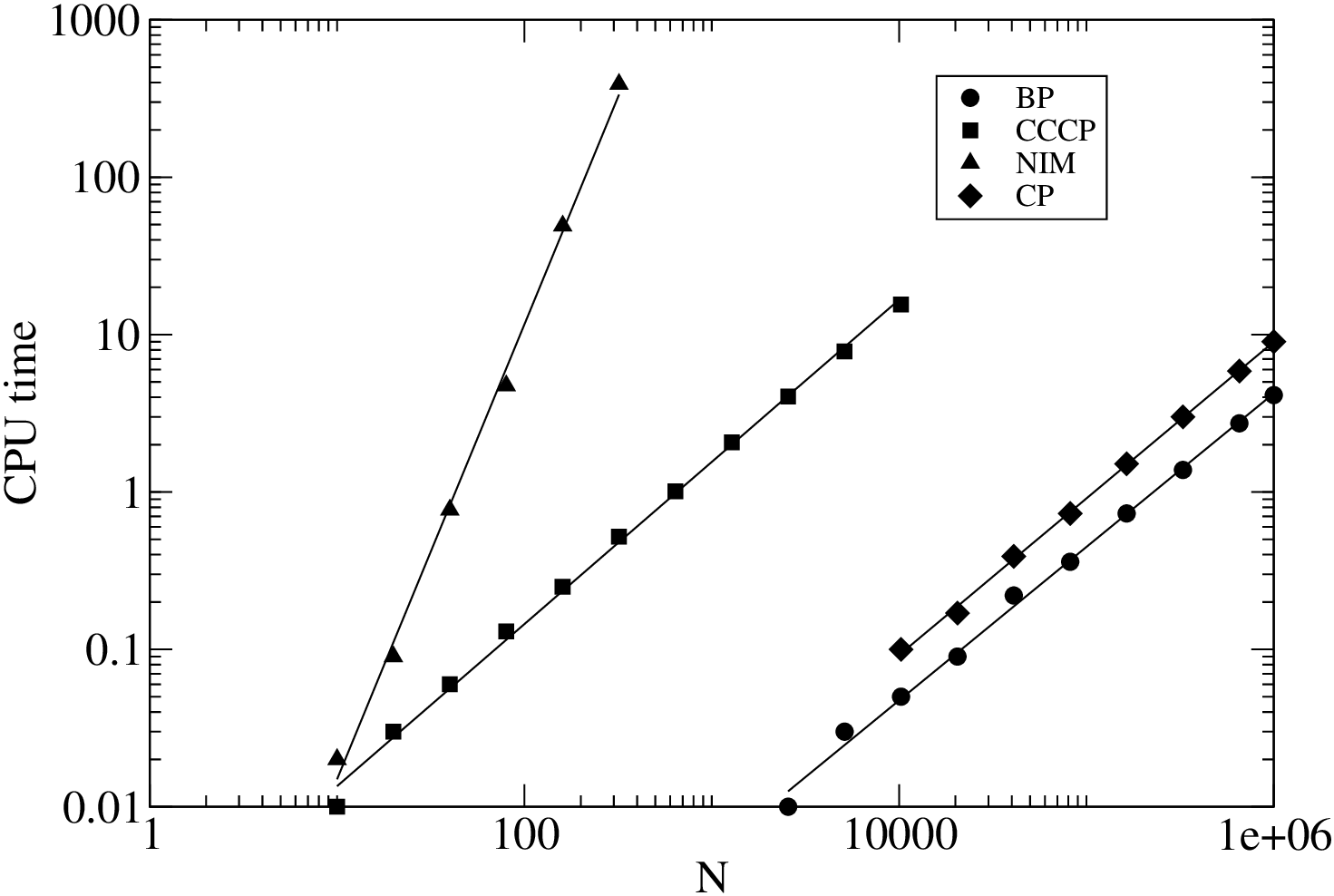}
\end{center}
\caption{\label{CPU1d}CPU times (seconds) for the 1d Ising chain with
random fields}
\end{figure}

Consider first a chain of $N$ Ising spins, with ferromagnetic
interactions $J>0$ and random bimodal fields $h_i$ independently drawn
from the distribution
\begin{equation}
p(h_i) = \frac{1}{2} \delta(h_i - h_0) + \frac{1}{2} \delta(h_i + h_0).
\end{equation}
The boundary conditions are open, and the model is exactly solved by
the CVM pair approximation. The various algorithms described are run
from a disordered, uncorrelated state and stopped when the distance
between two successive iterations, defined as the sum of the squared
variations of the messages (or the probabilities, or the Lagrange
multipliers, depending on the algorithm and the loop -- outer or inner
-- considered). \Fref{CPU1d} reports the CPU times obtained with
several algorithms, for the case $J = 0.1$, $h_0 = 1$. The HAK
algorithm is not reported since it reduces to BP due to the convexity
of the free energy. It is seen that the CPU time grows linearly with
$N$ for all algorithms except NIM, in which case it goes like
$N^3$. Despite the common linear behaviour, there are order of
magnitude differences between the various algorithms. While BP and CP
converges in 4 and 9 seconds respectively for $N = 10^6$, CCCP takes
15 seconds for $N = 10^4$. For NIM, finally, the fixed point is
reached in 12 seconds for $N = 10^2$. 

\begin{figure}
\begin{center}
\includegraphics*[scale=.5]{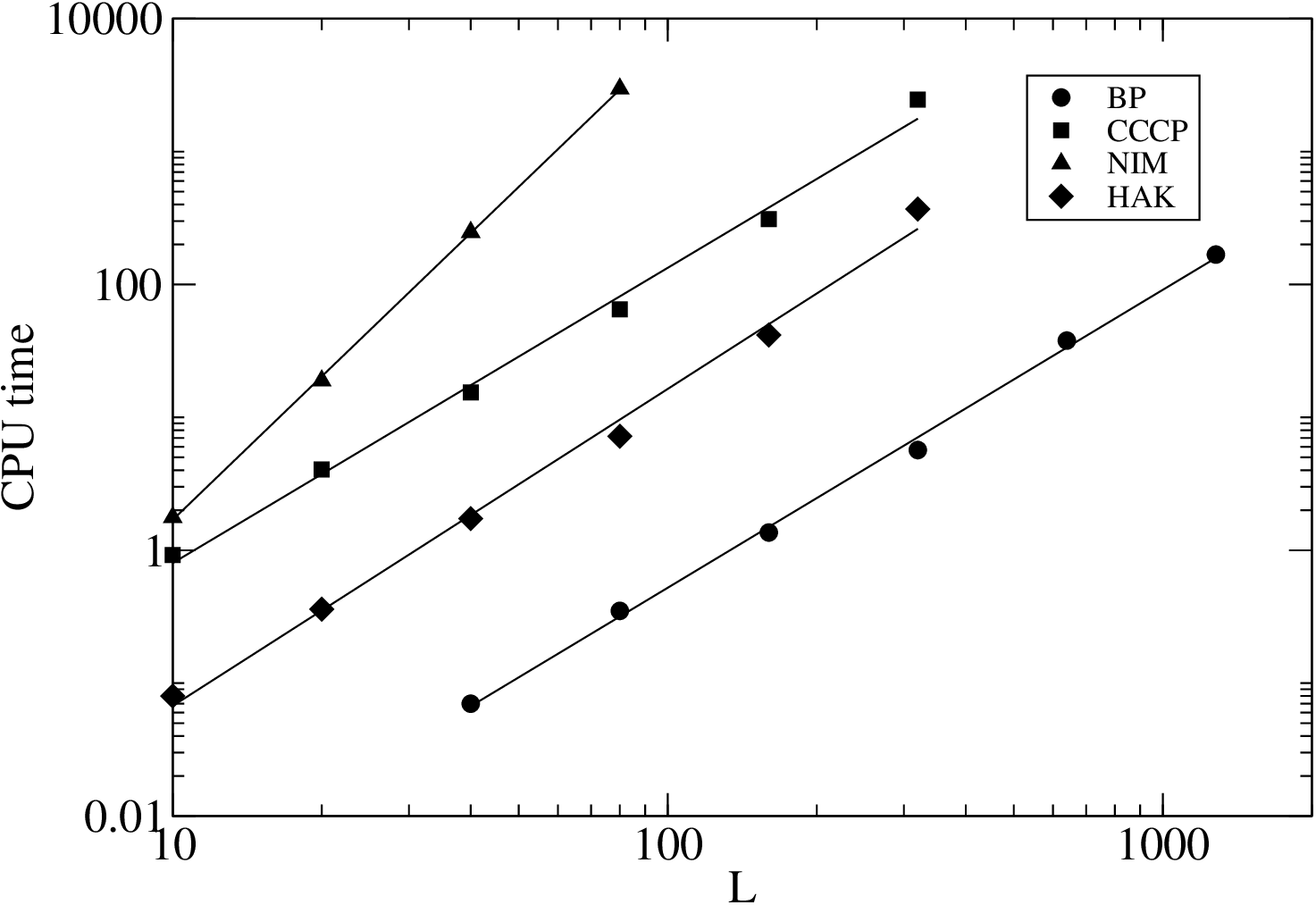}
\end{center}
\caption{\label{CPU2dP}CPU times (seconds) for the 2d
Edwards--Anderson model in the paramagnetic phase}
\end{figure}

As a further test, consider, again at the level of the pair
approximation, the two--dimensional Edwards--Anderson spin glass
model, defined by the Hamiltonian \Eref{Ising} with $h_i = 0$ and
random bimodal interactions $J_{ij}$ independently drawn from the
distribution
\begin{equation}
p(J_{ij}) = (1-p) \delta(J_{ij} - J) + p \, \delta(J_{ij} + J).
\end{equation}
Here the frustration effects are even more important and the
non--convergence problem of BP becomes evident. As a rule of
thumb, when the temperature, measured by $J^{-1}$, is small enough and
$p$ (the fraction of antiferromagnetic bonds) is large enough, the BP
algorithm stops converging. The condition for the instability of the
BP fixed point has been computed, in the average case, for Ising spin
glass models with pairwise interactions \cite{SG-BP-conv}. In order to
compare algorithm performances, \Fref{CPU2dP} reports CPU times vs $L$
for $N = L^2$ lattices with periodic boundary conditions, $J = 0.2$
and $p = 1/2$, that is well into the paramagnetic phase of the
model. The initial guess is a ferromagnetic state with $m_i = 0.9,
\forall i$. It is seen that the CPU times scale roughly as $N^{1.1}$
for all the algorithms considered except NIM, which goes like
$N^{1.8}$. Again the algorithms with linear behaviour are separated by
orders of magnitude. For $L = 320$ BP converges in 6 seconds, HAK in
370 seconds and CCCP in 2460 seconds.

CP has not been considered in the present and the following tests,
although empirically it is seen that its behaviour is rather close to
the HAK algorithm. Its performance is however severely limited as soon
as one considers variable with more than two states, due to a sum over
the configurations of the neighbourhood of a NN pair.

A similar comparison can be made in the ferromagnetic phase, setting
$J = 0.5$ and $p = 0.1$. Here the CPU times for the BP algorithm
exhibit large fluctuations for different realizations of the disorder,
and the data reported are obtained by averaging over 30 such
realizations. Now all algorithms exhibit comparable scaling
properties, with CPU times growing like $N^{1.5} \div N^{1.7}$. As far
as absolute values are concerned, for $L = 50$ convergence is reached
in 4, 44, 680 and 1535 seconds by BP, HAK, CCCP and NIM
respectively. 

\begin{figure}
\begin{center}
\includegraphics*[scale=.5]{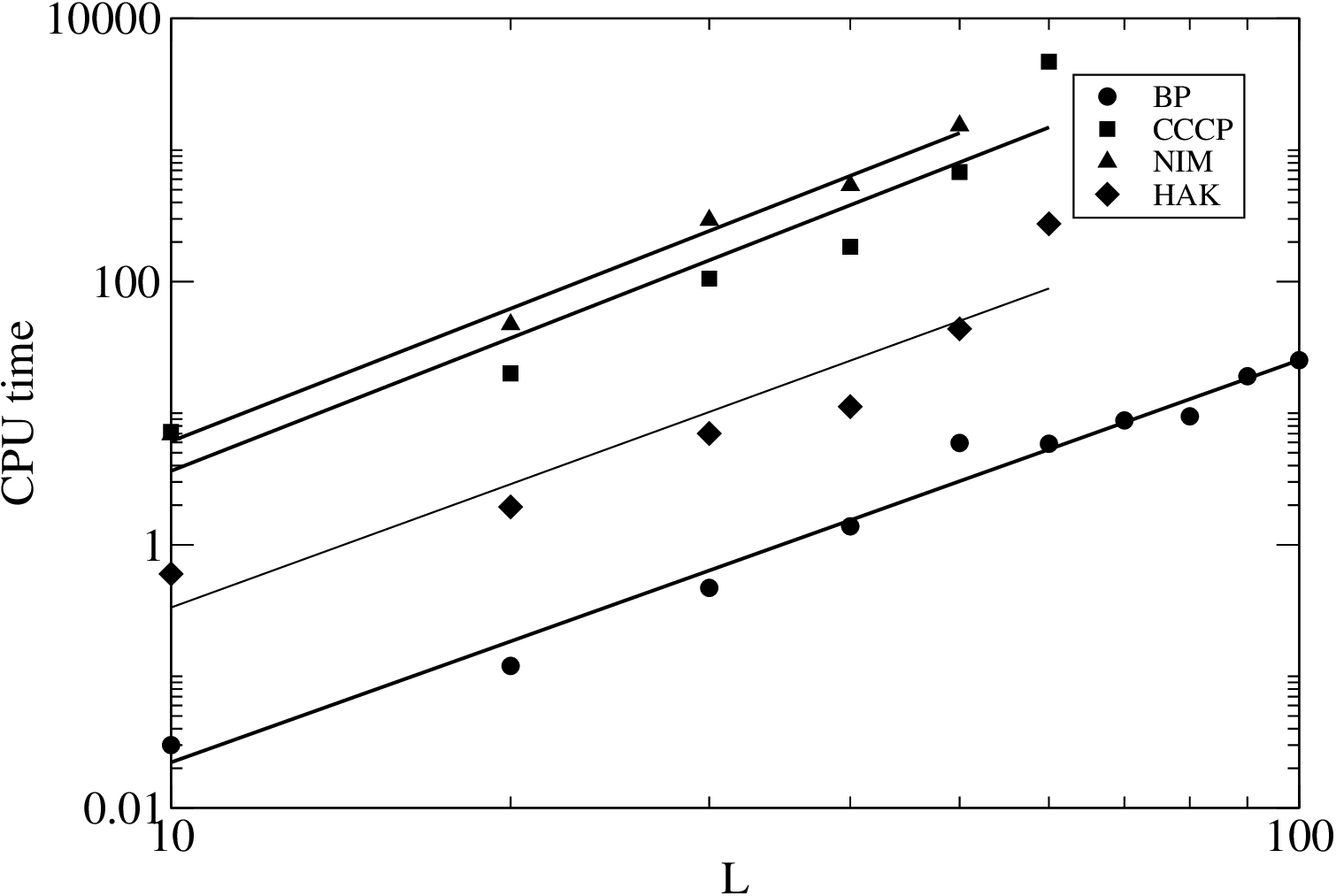}
\end{center}
\caption{\label{CPU2dF}CPU times (seconds) for the 2d
Edwards--Anderson model in the ferromagnetic phase}
\end{figure}

A similar scaling analysis was not possible into the glassy phase
(which is unphysically predicted by the pair approximation), due to
non--convergence of BP and too large fluctuations of the convergence
time of the other algorithms.

As a general remark we observe that BP is the fastest algorithm
available whenever it converges. Among the provably convergent
algorithms, the fastest one turns out to be HAK, at least in the
``just convex over the constraints'' \cite{HAK} scheme which was used
here. 

\section{Conclusions}
\label{Conclusions}

Some aspects of the cluster variation method have been briefly reviewed.
The emphasis was on recent developments, not yet covered by the 1994
special issue of Progress of Theoretical Supplement \cite{PTPS}, and
the focus was on the methodological aspects rather than on the
applications. 

The discussion has been based on what can be considered the modern
formulation of the CVM, due to An \cite{An88}, based on a truncation
of the cumulant expansion of the entropy in the variational principle
of equilibrium statistical mechanics. 

The advancements in this last decade were often due to the interaction
between two communities of researchers, working on statistical physics
and, in a broad sense, probabilistic graphical models for inference
and optimization problems. The interest of both communities is
currently on heterogeneous problems, while in the past the CVM was
most often applied to translation invariant lattice models (in this
topic, the only new advancements discussed have been the attempts to
extract information about critical behaviour from CVM results). The
more general point of view that has to be adopted in studying
heterogeneous problems has been crucial to achieve many of the results
discussed.

The formal properties of the CVM have been better understood by
comparing it with other region--based approximations, like the
junction graph method or the most general formulation of the
Bethe--Peierls approximation (the lowest order of the CVM), which can
treat also non--pairwise interactions. Studying realizability, that is
the possibility of reconstructing a global probability distribution
from the marginals predicted by the CVM, has led to the discovery of
non--tree--like models for which the CVM gives the exact solution.

A very important step was made by understanding that belief
propagation, a message--passing algorithm widely used in the
literature on probabilistic graphical models, has fixed points which
correspond to stationary points of the Bethe--Peierls
approximation. The belief propagation can thus be regarded as a
powerful algorithm to solve the CVM variational problem, that is to
find minima of the approximate free energy, at the Bethe--Peierls
level. This opened the way to the formulation of generalized belief
propagation algorithms, whose fixed points correspond to stationary
points of the CVM free energy, at higher level of approximation.

Belief propagation and generalized belief propagation are certainly
the fastest available algorithms for the minimization of the CVM free
energy, but they often fail to converge. Typically this happens when
the problems under consideration are sufficiently frustrated. In order
to overcome this difficulty double loop, provably convergent
algorithms have been devised, for which the free energy can be shown
to decrease at each iteration. These are similar in spirit to the old
natural iteration method by Kikuchi, but orders of magnitude faster,
though not as fast as BP and GBP.

When the frustration due to competitive interactions or constraints is
very strong, like in spin--glass models in the glassy phase or in
constraints satisfaction problems in the hard regime, even double loop
algorithms become useless, since we are faced with the problem of
replica symmetry breaking, corresponding to clustering of
solutions. Very important advancements have been made in recent years
by extending the belief propagation algorithm into this domain. These
results are in a sense at the border of the CVM, since they are at
present confined to the lowest order of the CVM approximation, that is
the Bethe--Peierls approximation. 

It will be of particular importance, in view of the applications to
hard optimization problems with non--tree--like structure, to
understand how to generalize these results to higher order
approximations.

\ack 

I warmly thank Pierpaolo Bruscolini, Carla Buzano and Marco Pretti,
with whom I have had the opportunity to collaborate and to exchange
ideas about the CVM, Marco Zamparo for a fruitful discussion about
\Eref{FactorProp}, Riccardo Zecchina for many discussions about the
survey propagation algorithm, and the organizers of the Lavin workshop
``Optimization and inference in machine learning and physics'' where I
had the opportunity to discuss an early version of this work.

\end{document}